\documentclass[prx,twocolumn,english,superscriptaddress,floatfix,longbibliography,nofootinbib]{revtex4-2}
\pdfoutput=1
\usepackage[utf8]{inputenc}
\usepackage[english]{babel}
\usepackage[T1]{fontenc}
\usepackage{color,graphicx,dsfont,slashed,subfigure,multirow,enumitem,amssymb,amsmath,hyperref,orcidlink}
\hypersetup{
	bookmarks=true,
	unicode=true,
	pdftoolbar=true,
	pdfmenubar=true,
	pdffitwindow=false,
	pdfstartview={FitH},
	pdftitle={Classical versus Quantum: comparing Tensor Network--based Quantum Circuits on LHC data},
	pdfauthor={J.Y.~Araz, M.~Spannowsky}, 
	pdfsubject={Quantum Machine Learning},
	pdfcreator={J.Y.~Araz, M.~Spannowsky},
	pdfproducer={texstudio}, 
	pdfkeywords={Quantum Machine Learning, Tensor Networks, Jets},
	pdfnewwindow=true,
	colorlinks=true,
	linkcolor=blue,
	citecolor=magenta,
	filecolor=magenta,
	urlcolor=cyan
}
\graphicspath{{figs/}}
\definecolor{grn}{RGB}{97,159,58}

\renewcommand{\figureautorefname}{Fig.}

\def\sectionautorefname~#1\null{Sec.~#1\null}
\def\subsectionautorefname~#1\null{Sec.~#1\null}
\def\figureautorefname~#1\null{Fig.~#1\null}
\def\tableautorefname~#1\null{Table~#1\null}
\def\equationautorefname~#1\null{Eq.~(#1)\null}

\begin{document}
	\title{Classical versus Quantum:\\ comparing Tensor Network--based Quantum Circuits on LHC data}

	\author{Jack~Y.~Araz\orcidlink{0000-0001-8721-8042}}
	\email{jack.araz@durham.ac.uk}
	\affiliation{Institute for Particle Physics Phenomenology,\\Durham University, South Road, Durham, DH1 3LE, UK}
	\author{Michael~Spannowsky\orcidlink{0000-0002-8362-0576}}
	\email{michael.spannowsky@durham.ac.uk}
	\affiliation{Institute for Particle Physics Phenomenology,\\Durham University, South Road, Durham, DH1 3LE, UK}
	
	\begin{abstract}
Tensor Networks (TN) are approximations of high-dimensional tensors designed to represent locally entangled quantum many-body systems efficiently. This study provides a comprehensive comparison between classical TNs and TN-inspired quantum circuits in the context of Machine Learning on highly complex, simulated Large Hadron Collider (LHC) data. We show that classical TNs require exponentially large bond dimensions and higher Hilbert-space mapping to perform comparably to their quantum counterparts. While such an expansion in the dimensionality allows better performance, we observe that, with increased dimensionality, classical TNs lead to a highly flat loss landscape, rendering the usage of gradient-based optimization methods highly challenging. Furthermore, by employing quantitative metrics, such as the Fisher information and effective dimensions, we show that classical TNs require a more extensive training sample to represent the data as efficiently as TN-inspired quantum circuits. We also engage with the idea of hybrid classical-quantum TNs and show possible architectures to employ a larger phase space from the data. We offer our results using three main TN ansatz: Tree Tensor Networks, Matrix Product States, and Multi-scale Entanglement Renormalisation Ansatz.
	\end{abstract}
	\maketitle
	\section{Introduction}\label{sec:intro}

	Machine Learning (ML) and Deep Neural Networks (DNN) have gained tremendous interest on the verge of technological developments and the availability of an abundance of data. The High Energy Physics (HEP) community have long been engaged with processing vast amounts of data generated by collider experiments like the Large Hadron Collider (LHC). Recent advancements in Neural Network (NN) technology allowed physics-driven analytic analysis methods to evolve into data-driven statistical approaches, which transformed the ability and accuracy of data analysis. The emergence of quantum computers has introduced another step of evolution in this formidable avenue. 
	
	Various quantum algorithms aim to tackle challenging tasks in optimisation problems and improve the interpretability of classical NNs applied to high-energy physics data. Such algorithms include but are not limited to simulations of collision events~\cite{hepsimu, helicityamp,Carena:2021ltu, Williams:2021lvr,li2021partonic, Bravo-Prieto:2021ehz}, reconstruction of the charged tracks~\cite{das2020track, qtrkreco, PhysRevD.105.076012, Tuysuz:2021tn}, and event classification analyses~\cite{Terashi:2020wfi, PhysRevResearch.4.013231, Wu:2020cye, Guan_2021, Blance:2020ktp, 2021andrew, Chen:2021ouz, Belis:2021zqi, Araz:2021un}. Despite the exceptional interest in quantum computation methods, there are still many open questions regarding the advantages of the quantum age~\cite{doi:10.1098/rspa.2017.0551}.
	
	There is particular interest in using the quantum paradigm in ML and optimization tasks, as the number of qubits needed is often already available in noisy intermediate-scale quantum (NISQ) devices, and error mitigation might be less of a concern. 
	Quantum NNs have various advantages over classical NNs \cite{Blance:2020nhl, Eisert2021,Roche2021, Abel:2021fpn, Ngairangbam:2021yma, PhysRevA.101.010301}, such as faster convergence and significantly better performance with the same network structure. Quantum algorithms achieve this by representing the correlations between input features within the quantum entanglement paradigm, providing a much richer data representation.
	
	 Tensor Networks (TN) are widely used to simulate strongly entangled quantum systems~\cite{Orus:2018dya, 10.5555/2011832.2011833}, and they can represent both quantum states and circuits~\cite{Shi_2006, Vidal_2008, Verstraete_2008}. Due to this property, TNs form a natural bridge between classical and quantum computation methods. 
	In the context of high-energy physics data analysis, Tensor Networks have recently been used for data originating from collider experiments. TN properties can be used to classify b-jets produced at the LHCb experiment where tabular data is analysed using Tree Tensor Network (TTN) architecture \cite{trenti2020quantuminspired}. Moreover, ref.~\cite{Araz:2021un} shows that Matrix Product States (MPS) can achieve state-of-the-art convolutional neural network (CNN) accuracies in top tagging via calorimeter images. By employing entanglement entropy information between MPS' tensor-blocks the same accuracy can be achieved with only 54\% of the pixels in a given calorimeter image.
	
	Given the ability to represent both NNs and quantum many-body systems, it is only natural to use TN-inspired quantum circuit architecture to transfer our classical knowledge on quantum hardware~\cite{Grant:2018vv, Huggins_2019, 10.3389/fphy.2020.586374, LAZZARIN2022128056, bhatia2019matrix, PhysRevResearch.2.033125, HUANG202189}. Besides the high accuracy rates, TN-based quantum circuits can lead to more robust results against noise in the near term quantum computers. Whilst this opens up an entirely new circuit design, near term quantum devices are still limited to a small number of qubits, restricting the usage of multimodal datasets. Hybrid classical-quantum TNs can be utilised to eliminate this issue to design end-to-end training sequences with classical data mapping and quantum classification layers~\cite{liu2020quantumclassical, chen2020hybrid}. This allows much larger datasets to be embedded in the optimisation process, and due to the duality of TNs, as more qubits are available, classical nodes can be transformed into quantum nodes to harness the full potential of the quantum hardware.
	
	This study investigates the usage of TN-based quantum variational circuits for top jet discrimination from the QCD background in calorimeter images. We have investigated three different architectures, namely Matrix Product States (MPS), Tree Tensor Networks (TTN) and Multi-scale Entanglement Renormalisation Ansatz (MERA) as quantum circuits and compared the results with their classical counterparts for one-dimensional data embedding. For simplicity, the quantum circuits with TN-inspired constraints on their gate structure will be referred to as quantum TN, and their source of inspiration will be classical TN\footnote{It is essential to note that this does not imply that TNs are classical or quantum, the naming choice has been made to simplify the flow of the study.}. Our results have shown that TNs require exponentially more trainable parameters with increasing qubit structure to achieve the same performances as their quantum counterparts, leading to computationally expensive network architectures for machine learning applications. We observed that the classical TNs require exponentially large bond dimensions to capture the same entanglement structure as the QTNs, leaving the stochastic gradient descent methods inefficient for optimising the tensor nodes. This has been improved by employing more extensive Hilbert space mapping of the input features, indicating that classical TNs require more information to represent the same data as well as QTNs. We present detailed numerical results to study quantum mechanical differences between TNs and QTNs. In the following, we investigate possible avenues for hybrid classical-quantum TN architectures, allowing a more extensive phase space to be used in the network.
	
	This study is organised as follows, in \autoref{sec:qtn_intro} we introduce TNs and QTNs as methods to perform machine learning tasks. The results have been discussed in \autoref{sec:results}, where we first introduce the dataset and preprocessing in \autoref{sec:preprocess} and then numerical analysis are presented in sections \ref{sec:tn_qtn} and \ref{sec:hybrid}.
	
	\section{Bridge between classical \& quantum Machine Learning}\label{sec:qtn_intro}
	Tensors are multidimensional objects that describe multilinear relations between algebraic entities defined on a particular vector space. This study employs ``tensor diagram notation'' to formalize the relation between tensors where a scalar is shown as a node (shown as a blue circle throughout this study), and each rank is shown with an external line to a given node~\cite{penrose1971applications, Orus:2013kga, Bridgeman:2016dhh}. 

	TNs are defined as a series of Einstein summations indicating the connections between tensor nodes, forming a graph of tensors. Unlike traditional graph networks, however, where each connection indicates the ``coupling strength'' between nodes, connections between TNs suggest the correlation between a node and the rest of the network and limits the range of entanglement between nodes. These connections between tensor nodes are called bond (or auxiliary) dimensions, shown as $ \chi $. Due to the computational cost of contraction of a TN, one can only form specific architectures of TNs that can be contracted efficiently. 

\begin{figure*}
		\centering
		\includegraphics[scale=.8]{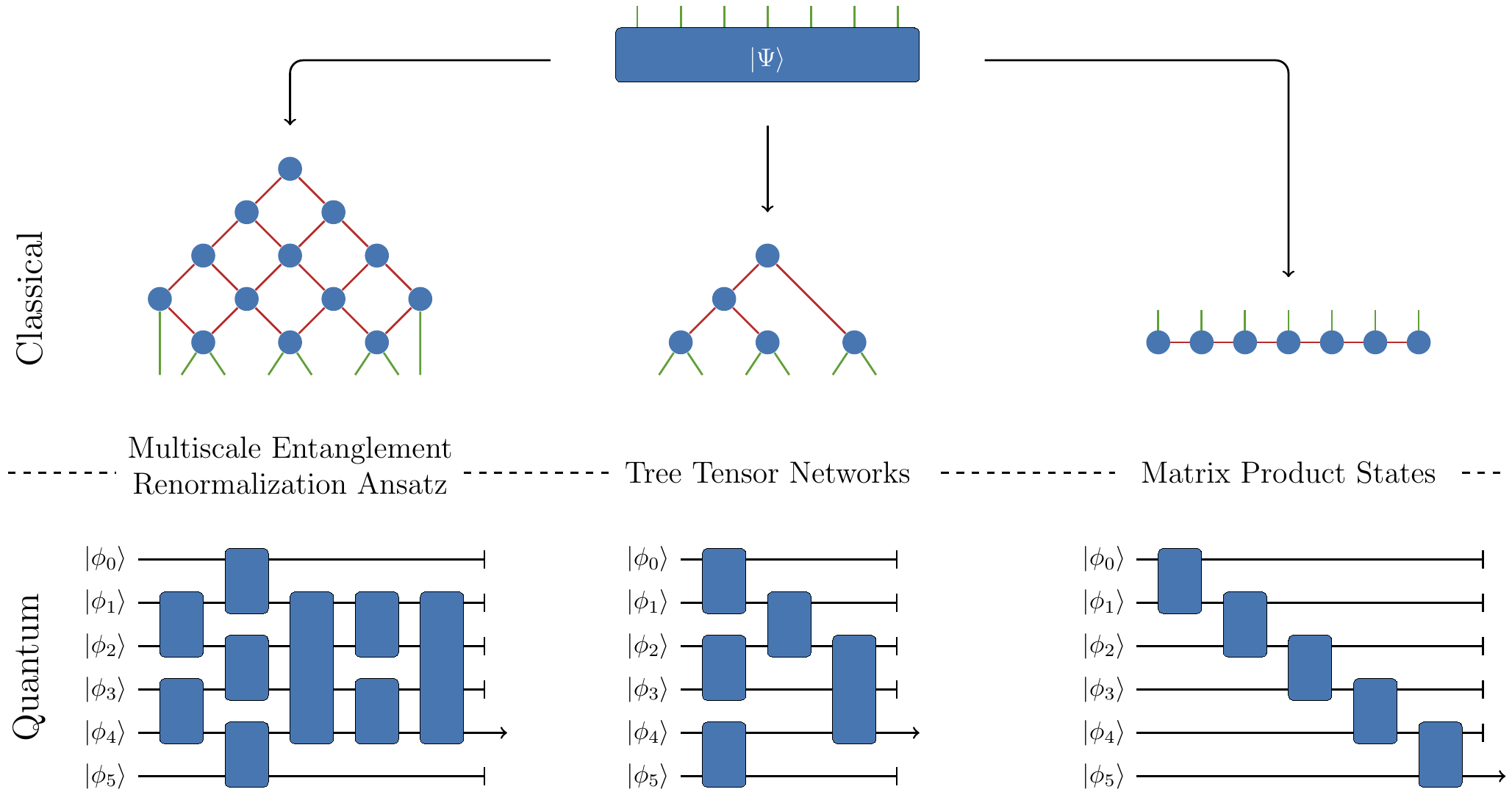}
		\caption{\it Upper panel shows the decomposition of the wave-function into various Classical Tensor Network realisations. Here the green edges represent Hilbert space dimensions, red lines represents auxiliary dimensions between tensors. The bottom panel shows their projection on a quantum circuits and each tensor block has been represented as certain transformation matrix between two qubits. Finally the arrow indicates a measurement. Images are only representative, varied number of inputs are given to simplify the interpretation of the network. \label{fig:qtns}}
	\end{figure*}

	As mentioned earlier, TNs mainly designed to study many-body quantum systems~\cite{Verstraete:2004cf} where one-dimensional lattice systems have been extensively studied, and many efficient architectures and contraction algorithms have been developed. A wave function for a one-dimensional lattice with $ N $ number of states can be written as
	\begin{eqnarray}
		| \Psi \rangle = \sum_{\phi_1, \dots, \phi_n \in \{0,1\}} \mathcal{W}_{\phi_1\cdots\phi_n} |\phi_1\rangle \otimes \cdots \otimes |\phi_n \rangle\ ,\label{eq:wavefunc}
	\end{eqnarray}
	where $ |\phi_i\rangle$ are spin states spanning a $ 2 $ dimensional local Hilbert space with $N$ state lattice. The subscript of each state identifies the state location within the lattice starting from the first state labelled as $1$ till the $n^{\rm th}$ state.  $ \mathcal{W}_{\phi_1\cdots\phi_n} $ is a rank-$ N $ amplitude tensor indicating the bond structure between each state.  Top of the Fig.~\ref{fig:qtns} shows $ | \Psi \rangle $ in tensor diagram notation where each Hilbert space dimension has been shown by green lines (for simplicity, only seven of them are shown on the image). Each green line represents a two-dimensional vector for a lattice of spin states where outer products form $ | \Psi \rangle $. The computational complexity of simulating such objects for spin states is $ \mathcal{O}(2^N) $, which grows exponentially with each additional state. However, it is possible to decompose this monstrous beast in smaller tensors that can efficiently represent the original tensor whilst reducing the computational cost of the object\footnote{Various tensor decomposition methods can be employed to achieve this, such as singular value decomposition~\cite{10.1093/qmath/11.1.50, Eckart:1936va}.}.
	
It is, perhaps, the most intuitive to decompose a one-dimensional lattice into MPS~\cite{Fannes:1992vq, Klumper:1992vi, doi:10.1137/090752286, Bridgeman:2016dhh, 10.5555/2011832.2011833, Orus:2013kga, Verstraete_2006, Hastings:2007iok, Chen:2010gda, Schollw_ck_2011}, where each state is entangled to the rest of the lattice through the adjacent states\footnote{For a detailed analysis on MPS in the context of HEP and machine learning, see ref.~\cite{Araz:2021un}.}. This allows a highly accurate approximation of $ |\Psi\rangle $ with only $ \mathcal{O}({\rm poly}(N)) $ parameters. Although this limits the entanglement range, MPS is a potent tool for simulating locally entangled states. The right branch of the classical portion of Fig.~\ref{fig:qtns} shows the representation of MPS in tensor diagram notation where each blue node represents a rank-3 tensor except for the ones on the edges, forming only rank-2 tensors. By utilising periodic boundary conditions, one can write a circular network where all nodes are rank-3 tensors. As mentioned before, green lines represent the Hilbert space dimensions where each blue tensor is connected to a state, $ |\phi_i\rangle$, on the lattice. Red lines between tensors represent the auxiliary dimensions, $ \chi $, where the size of $\chi$ determines the precision of the network by resizing the influence of each state to the rest of the lattice. Although a large bond dimension allows a more precise state representation, it also increases the computational cost of this representation. Hence, different architectures have been proposed to simulate more richly entangled lattice simulations.

Hierarchical TNs can be employed to capture a relatively more complex correlation structure. TTN is one of the widely used architectures in this avenue~\cite{Shi_2006}\footnote{For a detailed analysis on TTN in the context of HEP and machine learning, see ref.~\cite{trenti2020quantuminspired}.}. TTNs are constructed with feature condensing nodes where two or more vectors are condensed into a single vector. This allows neighbouring states to be mapped into a higher dimensional representation at each step. In machine learning (ML) terminology, each node can be seen as a local pooling layer. The middle branch of the classical portion of Fig.~\ref{fig:qtns} shows TTN in tensor diagram notation. Each green line shows the physical dimensions where neighbouring states are collected into one node and then mapped into a higher dimensional vector shown as red lines, the bond dimensions. Such a hierarchical structure allows each state to have more extensive entanglement allowing them to be entangled with further states on a higher-dimensional manifold. MPS, in this picture, can be classified as a maximally anti-symmetrised TTN.

	Further down the rabbit hole, Multi-scale Entanglement Renormalisation Ansatz (MERA) can be used to embed arbitrarily large entanglement structures into the network~\cite{Vidal_2008}. The left branch of Fig.~\ref{fig:qtns} shows MERA in tensor diagram notation. Although the figure shows a mixture of rank-4 transformation and rank-3 condenser nodes, MERA does not necessarily need condenser nodes; here, to simplify its application for classification, we mixed MERA with TTN. A different MERA architecture for classification can be found in ref.~\cite{reyes2020multiscale} where after several transformation layers with rank-4 tensors, authors reduced the dimensionality via an MPS layer. Each rank-4 tensor plays the role of transformation. In quantum field theory, such nodes have been employed to embed specific known symmetries into the network~\cite{Evenbly:2011tw}. Although such a network embodies much higher entanglement between states than MPS and TTN, the computational cost is much higher due to the loops that it forms within the architecture.

	Tensor Networks exhibit similarities with NNs~\cite{cohen2016convolutional, qentindl} where NN layers can be represented as TN layers, leading to more efficient and compressed network structures with similar or better outcomes~\cite{garipov2016ultimate}. TN architectures and quantum-inspired training algorithms can also be independently used as optimisation ansatz. In particular, specialised MPS training techniques can be used for classification tasks to achieve similar results to the state-of-the-art NN results~\cite{stoudenmire2017supervised, novikov2017exponential, selvan2020tensor, efthymiou2019tensornetwork, xu2021tensortrain}. 
	TTN has also been used for quantum assisted classification problems~\cite{PhysRevA.104.042408, liu2018machine, ttn2019} and MERA has been studied as symmetry embedding layer to a classifier~\cite{reyes2020multiscale, kong2021quantum}. Beyond optimisation problems, TNs can improve our understanding of training procedures and network structure due to their widely studied theoretical foundation~\cite{martyn2020entanglement}.

	For a given ML application, TNs require the feature space of a given data to be mapped onto a spin state lattice. An $ N $-dimensional feature space, $ \mathbf{x}\in\mathbb{R}^N $, can be written as outer products of $m$-dimensional vectors via a mapping function defined as $\phi(x):= \forall x\in\mathbb{R}\to\mathbb{C}^m$, where for a spin state $ m = 2$~\cite{stoudenmire2017supervised, Araz:2021un}. Hence the feature space takes the following form;
	\begin{align}
		\Phi^{\phi_1\cdots \phi_n}(\mathbf{x}) &= \bigotimes_{i=1}^N \phi_{i}(x_i)\quad  ;\label{eq:feature_map} \\ 
		\phi_{i}(x_i) &= \sum_{j=0}^{m-1} \alpha_j(x_i)|j\rangle\ .  \nonumber
	\end{align}
	Here each state is written as supper-position of spin states, $ \phi_{i}(x_i) \in \mathbb{C}^{m} $, with $ \alpha_j $ being feature dependent coefficients, forming an $N$-dimensional Hilbert space. The correlation between each state, in this form, is controlled by the amplitude tensor, $ \mathcal{W}_{\phi_1\cdots\phi_n} $, given in eqn. \eqref{eq:wavefunc}. The amplitude tensor represents the architecture of the TN, and depending on the decomposition sequence, it can be written as any TN architecture; for example, see Fig.~\ref{fig:qtns}.
	
	In a classification task, TNs produce the output probability 
	\begin{align}
		p\left(\mathbf{x}^{(i)};\theta\right) &= \left|f^l\left(\mathbf{x}^{(i)}\right)\right|^2 \nonumber \\ 
		&= \left| \mathcal{W}^l_{\ \phi_1\cdots\phi_n}(\theta)\ \Phi^{\phi_1\cdots \phi_n}(\mathbf{x}^{(i)})  \right|^2\ , \label{eq:cprop}
	\end{align}
	where $f^l(\mathbf{x}^{(i)})$ stands for the contraction of the network with the given $i$th data and $l$ represents the output label. $\mathcal{W}$ represented as a function of $\theta$, which are the trainable parameters of the network\footnote{Note that Einstein-summation convention has been employed throughout the paper.}. Traditional ML applications require each layer to be embedded into an activation function such as ReLu or sigmoid to capture the nonlinear properties of a given data. However, TNs can capture the nonlinearity without any activation function, ensuring an utterly linear network structure. $\mathcal{W}^l_{\ \phi_1\cdots\phi_n}$ consist of trainable parameters where each parameter, $\theta_j$, can be optimized with respect to a given loss function, $\mathcal{L}(\cdot,\cdot)$,
	\begin{eqnarray}
		\arg \min_{\theta_i\in\mathcal{W}}\ \mathcal{L} \left(q\left(\mathbf{x}^{(i)}\right),\ p\left(\mathbf{x}^{(i)}; \theta\right)\right)\ . \label{eq:opt}
	\end{eqnarray}
	Here $q\left(\mathbf{x}^{(i)}\right) = y^{\rm truth}$ represents the truth label of a given $ i $th data. The representation of the statistical data is the main topic of any ML application. 
	
	Entanglement entropy, so-called von Neumann entropy, can be employed as a metric to evaluate the expressiveness of TN states\footnote{For a review of entanglement entropy and area law in TNs see ref.~\cite{Eisert:2008ur} and for entanglement entropy in the context of quantum computing, see ref.~\cite{PhysRevResearch.2.033125}.}. For a bipartite system of TN, $ \mathcal{I} $, the entanglement entropy is defined as
	\begin{equation}
	 \mathcal{S}(\rho) = - {\rm Tr}[\rho \log_2 \rho]~, \nonumber
	 \end{equation}
	where $ \rho $ is the reduced density matrix of the system $ \mathcal{I} $. In a quantum many-body application a non-degenerate pure ground state, $\rho = |\Psi \rangle \langle \Psi|$, is expected to have vanishing entanglement entropy~\cite{10.5555/2011572.2011576, Plenio:2004he}. 
	The maximum entanglement entropy of an MPS, on the other hand, is limited by its bond dimension, $\chi$, hence limiting its maximum entanglement entropy to be proportional to $\mathcal{S}(\rho) \sim \ln\chi$. Thus, the required number of parameters in an MPS to represent a quantum system is exponentially large. 
	
	Similarly, MERA has been designed to have the intrinsic property of volume law where its maximum entanglement entropy is bounded by $\mathcal{S}(\rho) \sim s(\mathcal{I})\chi$. Here, $s(\mathcal{I})$ denotes the surface area of a TN graph~\cite{Eisert:2008ur}. Whilst both realisations provide an efficient approximation of a highly entangled system, each has a particular limitation that can significantly increase the computational complexity of the network for an ML application.

	Although the entanglement entropy represents the potential of a given ansatz in terms of representation of the underlying data, the trainability of a given model is also an important aspect. In order to efficiently use gradient-based methods, the optimization landscape is required not to be flat. One of the biggest challenges of QML applications are barren plateau where the gradient of the loss function is exponentially suppressed; hence it is highly challenging to train the model using gradient-based methods~\cite{McClean:2018um}. However, the Fisher information matrix can be used to quantify the trainability of a given ansatz which shows the information gained by a given parametrization ansatz~\cite{martens2020new, berezniuk2020scaledependent, Abbas:2021wp}. For a given probability distribution in eqn.~\eqref{eq:cprop}, the mean Fisher information is calculated as the variance of the partial derivative with respect to the model parameters, $\theta$, of the log-likelihood
	\begin{eqnarray}
		 \bar F(\theta)  &=& \frac{1}{|\mathbf{X}|} \sum_{x\in\mathbf{X}} \partial_\theta\log p(x;\theta)\ \partial_\theta \log p^\top(x; \theta)\ , \nonumber 
	\end{eqnarray}
	where $\mathbf{X}$ is the training sample set. $\bar F(\theta)\in\mathbb{R}^{d\times d}$, for $d$-parameters, forms a Riemannian metric capturing the sensitivity of the ansatz with respect to the change in the parameters. In a classical network, the eigenvalue distribution of the normalised $ \bar F(\theta) $ is mostly degenerate around zero with rare large values~\cite{pmlr-v89-karakida19a}. Such distribution shows that the ansatz is not sensitive to the change of most of the parameters.

	Based on Fisher information, ref.~\cite{Abbas:2021wp} has introduced a measure of normalised effective dimensions
	\begin{eqnarray}
		\hat{d}_{\rm eff} = \frac{2\log \left( \frac{1}{V_\Theta} \int_{\Theta} \sqrt{ \det \left( \mathds{1} + \frac{|\mathbf{X}|}{2\pi\log|\mathbf{X}|}\hat{F}(\theta) \right) } d\theta \right)}{d \log\left( \frac{|\mathbf{X}|}{2\pi\log|\mathbf{X}|} \right)}\ , \nonumber
	\end{eqnarray}
	where $\hat{F}(\theta)$ is normalised Fisher information matrix. $V_\Theta$ is the volume of the parameter space $\theta_i \in \Theta$. At the $|\mathbf{X}|\to\infty$ limit, $\hat{d}_{\rm eff}$ of any given ansatz converges to one, but the convergence of an ansatz is slowed down by small and uneven eigenvalues of normalised Fisher matrix.

\medskip 

	A quantum algorithm is designed through a network of quantum gates to compute specific tasks. The availability of Quantum Machine Learning (QML) methods triggers the question of whether there is a viable quantum gate architecture to harness quantum advantage for ML applications. TNs can be represented as quantum circuits and hence can pose as a viable option for a variational quantum circuit (VQC)~\cite{Huggins_2019}. Due to the theory built to understand the training and the network structure of TNs, it poses as a powerful variational circuit option for ML applications~\cite{Grant:2018vv}.
	
	\begin{figure}
		\centering
		\includegraphics[scale=1.]{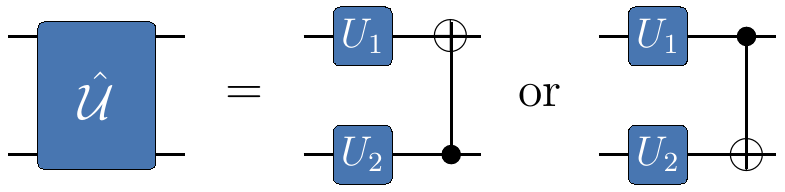}
		\caption{\it Representation of a tensor block in a quantum circuit. Each tensor block involves a unitary transformation, $ U(\theta, \varphi, \lambda) $, followed by a CNOT gate.\label{fig:tensor_block}}
	\end{figure}

	An MPS-inspired quantum variational circuit (Q-MPS) can be written by applying a set of unitary transformations to the initial adjacent two-qubit system. Each following two-qubit transformation block takes the last output qubit from a previous block and entangles it to the following qubit. For a six-qubit input, a Q-MPS construction is shown on the right branch of the bottom panel of the Fig.~\ref{fig:qtns}. Each blue transformation block represents two unitary transformations, as shown in Fig.~\ref{fig:tensor_block}, followed by a CNOT gate to entangle two states where a generic unitary transformation and the CNOT gate are expressed as 
	\begin{widetext}
	\begin{eqnarray}
		U(\theta, \varphi, \lambda) &= \left( \begin{array}{ccc}
			\cos(\theta/2) & & -e^{i\lambda}\sin(\theta/2) \\
			e^{i\varphi}\sin(\theta/2)& & e^{i(\lambda+\varphi)}\cos(\theta/2)\\
		\end{array}\right)\quad ,\quad
	{\rm CNOT} &= \left( 
		\begin{array}{cccc}
			1 & 0 & 0 & 0 \\
			0 & 1 & 0 & 0 \\
			0 & 0 & 0 & 1 \\
			0 & 0 & 1 & 0 \\
		\end{array} \right)\ . \label{eq:u_cnot}
	\end{eqnarray}
	\end{widetext}
	Here $ \theta,\ \varphi $ and $ \lambda $ are trainable variables. This definition forms the minimal Q-MPS construction. The block structure can also be enhanced via multiple CNOT gates, or auxiliary qubits can be introduced between circuit blocks to increase the bond dimension. Additionally, blocks allowing more qubits has also been proposed in other studies ({\it e.g.} see ref.~\cite{Huggins_2019}). To get the classification output, one needs to measure the expectation value of the last entangled qubit,
	\begin{widetext}
		\begin{eqnarray}
			\mathcal{M}_{\mathbf{\theta}} (\Phi^{\phi_1\cdots \phi_n}(\mathbf{x}) ) = \langle \Phi |\ \mathcal{\hat{U}}^\dagger_{\rm QC}( U_i(\theta_j) )\ \mathcal{\hat{M}}\ \mathcal{\hat{U}}_{\rm QC}( U_i(\theta_j) )\   | \Phi \rangle\ , \label{eq:measurement}
		\end{eqnarray}
	\end{widetext}
	where $ \mathcal{\hat{U}}_{\rm QC}( U_i(\theta_j) ) $ is the given quantum circuit (QC) constructed by unitarities, $U_i(\theta_j)$, with a set of free parameters, $\theta_j$, and $\mathcal{\hat{M}}$ is a single-qubit operator which we will choose as the third Pauli matrix, $\mathcal{\hat{M}} = \hat{\sigma}_z$. As before, for a classification task, one can define the probability of the network output as the mod square of the expectation value defined in eq.~\eqref{eq:measurement}, 
	\begin{eqnarray}
		p\left(\mathbf{x}^{(i)}; \theta\right) = \left| \mathcal{M}_{\mathbf{\theta}} \left(\Phi^{\phi_1\cdots \phi_n}\left(\mathbf{x}^{(i)}\right) \right) \right|^2\ , \label{eq:qprop}
	\end{eqnarray}
	and trainable parameters of the system, $ \theta $, can be optimised by minimising eqn. \eqref{eq:opt}.

	Similarly, a TTN-inspired quantum variational circuit (Q-TTN) can be constructed by applying a unitary block to each set of adjacent qubit. One can then discard one of the qubits and apply another unitary block to each remaining adjacent qubit. By repeating this process, one can connect the qubits in a hierarchical structure to perform a measure on the top-level qubit. The middle branch of the bottom panel of Fig.~\ref{fig:qtns} shows a six-qubit representation of Q-TTN.
	
	The MERA-inspired quantum variational circuit (Q-MERA) is closely related to Q-TTN, but a set of additional unitaries has enhanced the network ahead of each Q-TTN layer. These additional layers allow the circuit to capture more enhanced correlations between qubits that have a similar role in classical MERA.
	
	Each of the QC representations given in Fig.~\ref{fig:qtns} show an initial state of $|\phi_i\rangle^{\otimes N}$. However, the initial state of quantum hardware is $|0\rangle^{\otimes N}$; hence a data encoding process has to take place. This can be achieved in two ways; the data can be encoded in each individual qubit amplitude (qubit encoding) or by encoding the data on the entangled state amplitude (amplitude encoding)~\cite{MLQC}. In this study, we will use qubit encoding by rotating each qubit on the y-axis with respect to the input values where the rotation is defined as $R_y(x) = U(x,0,0)$, as defined in eqn. \eqref{eq:u_cnot}.

	In the following section, we will demonstrate the usage of these architectures in the context of top tagging against QCD background. We have proposed two types of network ansatz where first, we will employ purely TN-inspired VQC, and then we will introduce a hybrid ansatz where the data is processed by a classical TN before passing it to the VQC for classification.

	\section{Top tagging through Quantum Tensor Networks}\label{sec:results}
	
	The nature of electroweak symmetry breaking (EWSB) has yet to unfold. Due to the sizeable corresponding production cross-sections, currently, there are billions of top quarks produced at the LHC. With improved capabilities at high luminosity and high energy LHC in the coming years, the top physics will move to an even higher differential precision era. The large mass of the top quark gives it a unique property of high coupling to the Higgs boson. Accessing the top quark's properties and coupling strength can bring us closer to understanding the nature of EWSB. However, the measurement of top quark's properties has been obscured mainly due to high levels of background originating from QCD radiation, limiting the analyses to relatively cleaner leptonic final states of the top production. This significantly reduces the production cross-section hence the sensitivity to its properties. In turn, this led the HEP community to investigate the internal structure of jets collimated sprays of radiation~\cite{Marzani:2019hun}, where various analytical reconstruction techniques have been developed to understand the substructure of jets originating from different particles. This enabled a sophisticated, precise theoretical understanding of the plethora of highly complex data which can not be found in any field of science. With the dawn of the deep learning era, these attempts have been shifted towards data-driven analyses.
		
	At the LHC, particularly ATLAS and CMS experiments, the hadronic decay channels of the top quark have been widely investigated. The so-called jet objects, including the information for hadronic activity, deposit their energy in the electronic and hadronic calorimeters in these experiments. These calorimeters can be interpreted as a pixellated cylindrical camera recording the transverse momentum of the radiation shower. 
	
	As a demonstration of the capabilities of these networks, we choose to study the discrimination of top quarks against QCD background images reconstructed from energy deposits of the constituents in the electromagnetic and hadronic calorimeters in ATLAS experiment. Calorimeter images provide a natural correlation between pixels, making entangled states highly advantageous for classification processes. Additionally, CNN architectures built to classify jet images have been shown to be highly successful in discriminating gluons from quarks or tops from QCD backgrounds. In this study, we use these images by mapping them onto a relevant form for the given Tensor Network (quantum and classical) to process and classify. The code implementation of this study can be found at \href{https://gitlab.com/jackaraz/tnqcircuits}{this link}\footnote{\href{https://gitlab.com/jackaraz/tnqcircuits}{https://gitlab.com/jackaraz/tnqcircuits}}.

	\subsection{Dataset \& Preprocessing}\label{sec:preprocess}

	In this study, the usage of QTN classifiers has been demonstrated via the dataset provided in~\cite{Kasieczka:2019dbj, kasieczka_gregor_2019_2603256}, often used for benchmarking classification algorithms. It contains collider events for hadronic tops and QCD jets at $\sqrt{s} = 14$ TeV. Following the event generation and parton shower in \textsc{Pythia}~8~\cite{Sjostrand:2014zea}. The detector simulation has been implemented by means of \textsc{Delphes}~3 package~\cite{deFavereau:2013fsa} using its default ATLAS configuration card. In order to capture collimated boosted top topology, jets are reconstructed via \texttt{anti-kT} algorithm~\cite{Cacciari:2008gp} with $R=0.8$ embedded in \textsc{FastJet}~\cite{Cacciari:2011ma} package. These jets are also required to be within $p_T \in [550, 650]$ GeV and bounded by $|\eta|<2$. Top jets are tagged via parton matching using jet radius as the boundary for the angular separation between the jet and the parton level top quark. The dataset includes 1.2 million training and 400,000 separate test and validation samples, respectively.

	\begin{figure*}[!t]
		\centering
		\includegraphics[scale=.32]{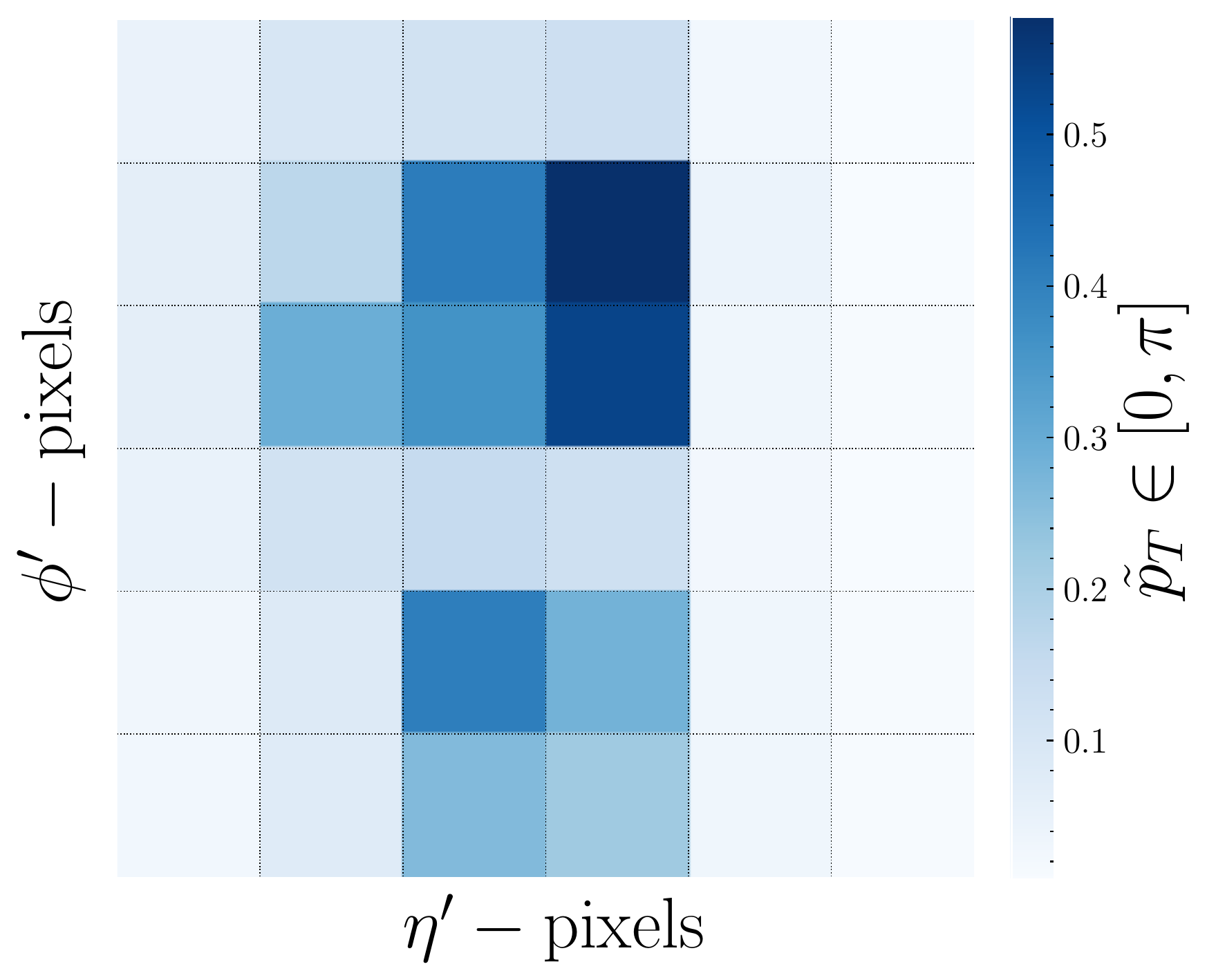}
		\includegraphics[scale=.32]{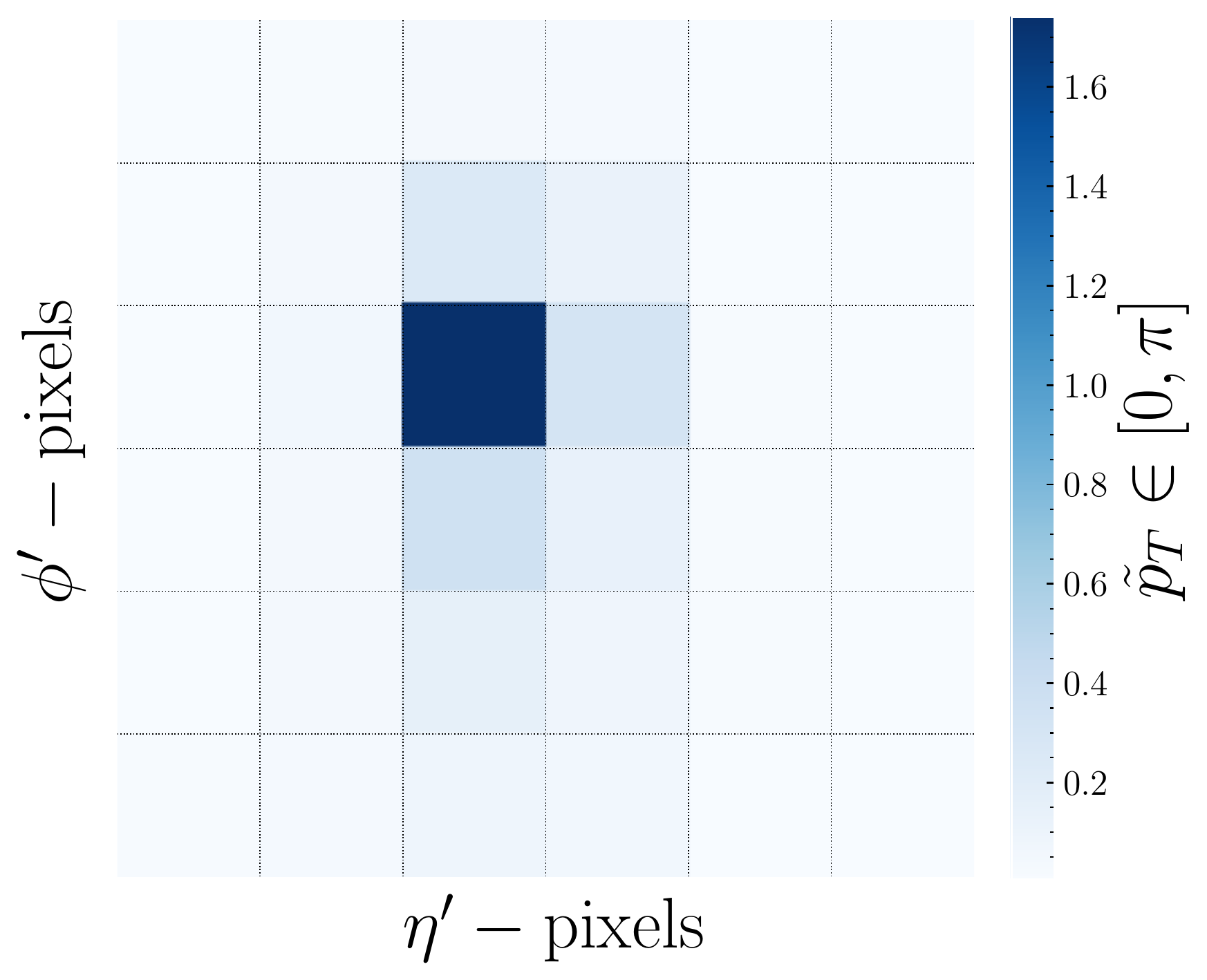}
		\includegraphics[scale=.32]{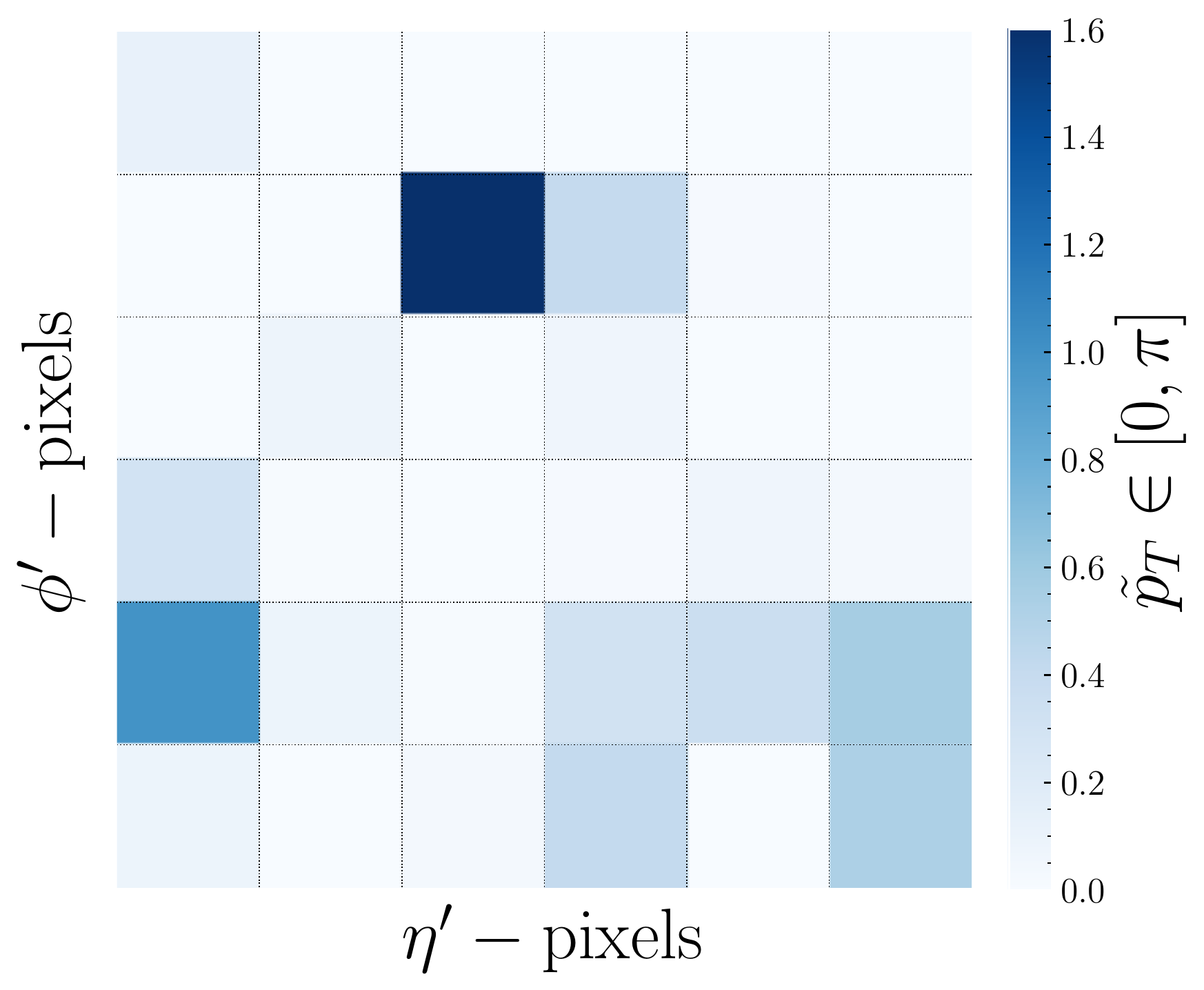}
		\caption{\it The left panel shows an average of 5000 top events, and similarly middle panel shows the same for the QCD background. The right panel shows the top signal for a single event. All images are reduced by cropping twelve pixels from each side and downsampling by averaging four adjacent pixels into one. Colour bar shows the intensity of the modified $ p_T $ deposited in each reduced pixel which has been fitted within $ [0,\pi] $. \label{fig:sig_bkg_im}}
	\end{figure*}

The jet images are prepared and standardized by following the prescription given in refs.~\cite{Araz:2021wp, Araz:2021un} where the constituents of the leading jet, defined above, has been centred on a $p_T$ weighted centroid in $\eta-\phi$ - plane. All the modified constituents of the leading jets have been mapped into a $37\times 37$ pixelated frame in the $\eta-\phi$ - plane spanning within $[-1.5, 1.5]$ range. Finally, the most energetic image quadrant has been moved to the top right by horizontally and vertically flipping the image. All training samples have been standardized by fitting the pixel values within $[0,\pi]$ range for 200,000 mixed-signal and background samples. Fig.~\ref{fig:sig_bkg_im} shows the standardized average of 5000 events for top signal (left panel) and QCD background (middle panel). Additionally, the right panel presents a single top signal event. The image has been prepared by cropping twelve pixels from each side and downsampling it by averaging four pixels into one, reducing the image size to $6\times 6$ without losing any vital information.

	In order to feed the data into the classical and quantum TNs, the data has to be processed a step further. Since each initial state in a quantum circuit is $|0\rangle$, one needs to prepare this initial state to represent the given data point. Whilst there are various ways of mapping the data on a quantum circuit (see ref.~\cite{MLQC}), we chose to map it by rotating each state around the y-axis by the corresponding pixel value using $R_y(\tilde{p}^i_T) = U(\tilde{p}^i_T, 0, 0)$ with $U(\theta, \phi, \lambda)$ defined in Eqn. \eqref{eq:u_cnot}. Similarly, as shown in ref.~\cite{stoudenmire2017supervised}, classical TNs require $ D $-dimensional mapping of the data to simulate the Hilbert space. Such mapping can be performed by using
	\begin{widetext}
	\begin{eqnarray}
		\phi^{i}(\tilde{p}^i_T) = \sqrt{ 
			 \begin{pmatrix}
				D-1 \\ 
				d_j - 1
			\end{pmatrix} }\ \cos^{D-d_j}\left(\tilde{p}^i_T \frac{\pi}{2}\right)\ \sin^{d_j - 1} \left(\tilde{p}^i_T \frac{\pi}{2} \right)\ ,\quad d_j \in \{1,\ldots, D\}\ ,\label{eq:hypersphere}
	\end{eqnarray}
	\end{widetext}
	where for $ D=2 $, $\phi^{i}(\tilde{p}^i_T) $ reduces to $ [\cos(\tilde{p}^i_T \pi /2),\ \sin(\tilde{p}^i_T \pi /2)]$.

	\subsection{Network architecture \& training}\label{sec:training}

	In this section, we will demonstrate a comprehensive numerical analysis and comparison between QTNs and TNs with different sized networks and architectures along with possible hybrid implementations. Our framework relies on \textsc{TensorFlow} (version~2.7.0)~\cite{tensorflow2015-whitepaper, DBLP:journals/corr/AbadiBCCDDDGIIK16} where quantum circuits are simulated using \textsc{PennyLane} (version~0.20.0)~\cite{bergholm2020pennylane} with \textsc{Qiskit} (version~0.32.1) backend~\cite{Qiskit} (\textsc{PennyLane--Qiskit} plugin version 0.20.0).

	\subsubsection{Classical vs. Quantum Tensor Networks}\label{sec:tn_qtn}

To compare the usage of different architectures presented in the previous section, we prepared two sets of ansatz, one for four qubits and another for six qubits. Such small feature space is owed to the currently accessible quantum hardware limitations, here IBM's quantum hardware. Hence, to employ an extended feature space, we also provide a hybrid classical-quantum TN where the classical portion of the network is responsible for mapping the image into a lower-dimensional manifold, which can then be deposited into the quantum hardware.

	To limit the feature space for four and six qubits, the calorimeter image of $37\times37$ pixels, presented in \autoref{sec:preprocess}, has been cropped from each side by fourteen pixels and then downsampled, leaving an image of $4\times4$ pixels. We used central four pixels for four qubit networks. For six-pixel networks, we added the adjacent two top pixels of the image. As prescribed in \autoref{sec:preprocess}, the modified transverse momentum in each of these pixels is then fitted within $[0,\pi]$. In the following, the image has been reshaped to form a vector.
	\begin{figure*}
		\centering
		\includegraphics[scale=1.25]{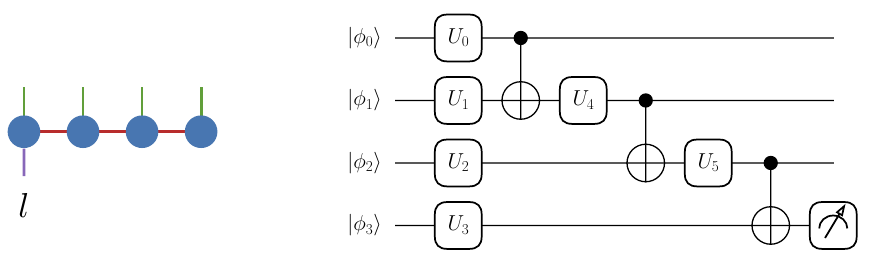}
		\caption{\it Representation of the four qubit classical MPS on the left and Q-MPS on the right.\label{fig:mps_nq4}}
	\end{figure*}
	
\begin{figure*}
		\centering
		\includegraphics[scale=1.5]{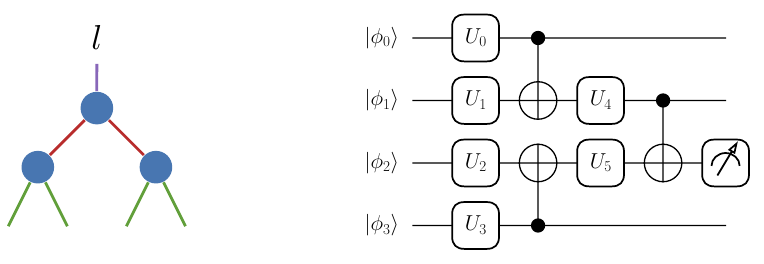}
		\caption{\it Representation of the four qubit classical TTN on the left and Q-TTN on the right.  \label{fig:ttn_nq4}}
	\end{figure*}

\begin{figure*}
	\centering
	\includegraphics[scale=1.25]{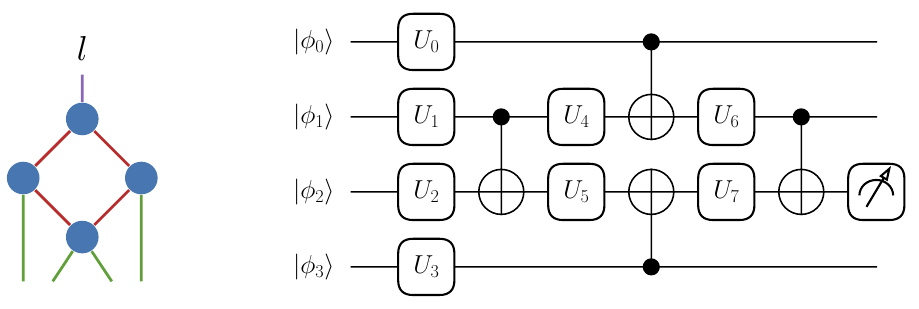}
	\caption{\it Representation of the four qubit classical MERA on the left and Q-MERA on the right.   \label{fig:mera_nq4}}
	\end{figure*}

	Figures \ref{fig:mps_nq4}, \ref{fig:ttn_nq4} and \ref{fig:mera_nq4} shows the four qubit construction of classical TNs (on the left) and Q-TNs (on the right) for MPS, TTN and MERA respectively. Each TN is required to have bond dimensions of $\chi=5$ (shown via red lines), and the physical dimensions (shown via green lines) are set to $D=2$ via Eqn. \eqref{eq:hypersphere}. Note that both bond and physical dimensions of the TNs can be set to much larger values, but we will limit them to compare with the QTN performance. The purple line in each TN shows the prediction output.

	Each QTN has been constructed via a set of unitary blocks, which includes trainable parameters where we will limit each unitary block to $U_i(\theta_i,0,0)$ by fixing $\varphi $ and $ \lambda$ to zero. As seen in figures \ref{fig:mps_nq4} and \ref{fig:ttn_nq4} both MPS and TTN network possesses six unitary transformation giving each six trainable parameters. Q-MERA, shown in Fig.~\ref{fig:mera_nq4}, on the other hand, possesses eight trainable parameters.

\begin{figure*}
		\centering
		\includegraphics[scale=1.5]{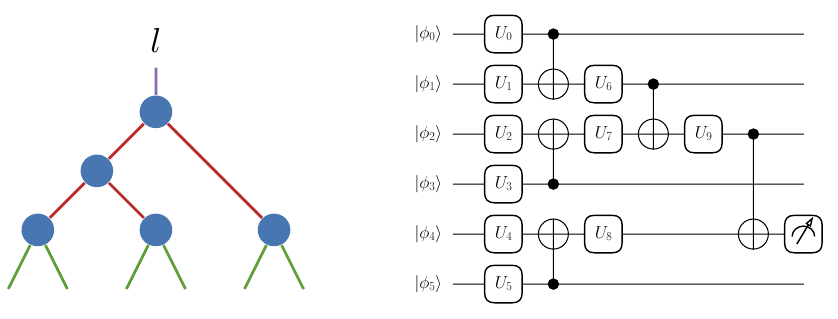}
		\caption{\it Representation of the six qubit classical TTN on the left and Q-TTN on the right.  \label{fig:ttn_nq6}}
	\end{figure*}

\begin{figure*}
		\centering
		\includegraphics[scale=1.28]{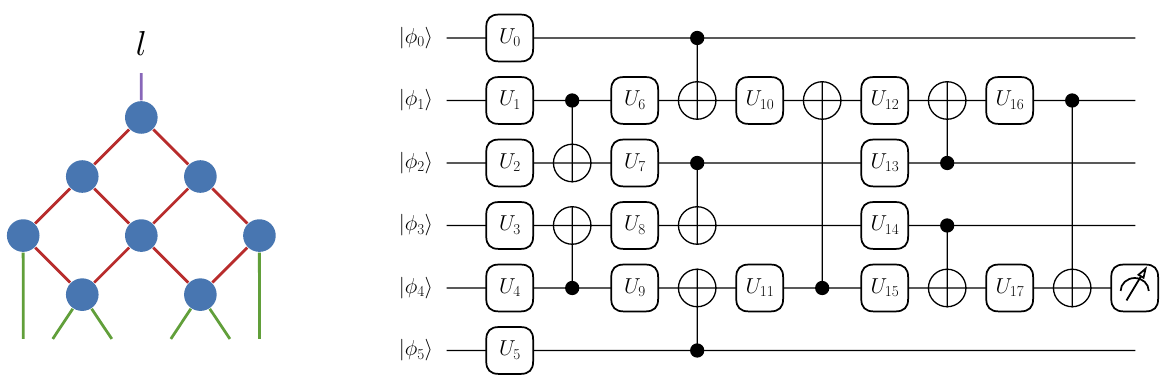}
		\caption{\it Representation of the six qubit classical MERA on the left and Q-MERA on the right.  \label{fig:mera_nq6}}
	\end{figure*}

	Similarly, figures \ref{fig:ttn_nq6} and \ref{fig:mera_nq6} shows the six qubit configuration for TNs (on the left) and QTNs (on the right) for TTN and MERA, respectively. Due to its monotonic architecture, an image for MPS hasn't been included, but the six-qubit version can be visualized by adding two more qubits to Fig.~\ref{fig:mps_nq4} with the same pattern. Q-MPS and Q-TTN have nine trainable parameters in this configuration, whereas Q-MERA possesses seventeen trainable parameters due to its complex structure.
	
	For the training of TNs, we employed \texttt{Adam}~\cite{Kingma2014AdamAM} optimization algorithm with a learning rate of $10^{-4}$. Although it has been noted in a previous study that training TNs with normalized gradients leads to much more stable tensor evolution~\cite{Araz:2021un}, we will employ standard gradient descent for the TN training. For four qubit sample, we used a batch size of 100 events; however, for the six-qubit configuration, this batch size led unstable training sequence hence reduced to 50 events per batch.

	For QTNs, we employed Quantum Natural Gradient Descent (QNGD)~\cite{rebentrost2018quantum, qng2020}, which improves the convergence speed in the training of variational quantum circuits \cite{Blance:2020nhl}, for faster optimization with a learning rate of $10^{-2}$. Instead of directly updating the given trainable parameters via their gradients for a given loss function, this algorithm solves a linear equation, $ \mathbf{M}\lambda = \mathbf{G}$ for $ \lambda $, where $\mathbf{M}$ is the metric tensor of a given circuit, and $\mathbf{G}$ is the gradient tensor. Then the trainable parameters are updated via $\tilde{\theta}_i = \theta_i - \eta \lambda_i$, where $\theta_i$ are the trainable parameters and $\eta$ is the learning rate. The training has been limited to 100 events per batch, and both classical and quantum ansatz has been trained with the full training sample.

	Each ansatz has been trained with the cross-entropy loss function, 
	\begin{eqnarray}
		\mathcal{L} = - \frac{1}{N}\sum_{x\in\mathbf{X}} y^{\rm truth}\ \log \left(p\left(x; \theta\right) \right) \ , \label{eq:loss}
	\end{eqnarray}
	where both TNs and QTNs are assumed to be Born machine; hence $ p\left(x; \theta\right)  $ given in equations \eqref{eq:cprop} and \eqref{eq:qprop}, respectively. During the training of each ansatz, we observed that loss value evolution for TNs is much slower than QTNs to converge to a minimum loss value; hence TNs are trained for $ 200 $ epochs, where QTNs are only allowed to train for $ 100 $ epochs. We required the training to be terminated if there is no improvement in validation loss value for $ 50 $ iterations, resulting in QTNs only training for $ 50-55 $ epochs depending on the ansatz; however, all TN ansatz has run for the entire $ 200 $ epochs. We find that QTN training could be terminated within the first $ 20-25 $ epochs for both four and six-qubit configurations. In addition to the early termination condition, we also required learning rate decay for every $ 25 $ epoch by a factor of $ 0.5 $, depending on the improvement of the validation loss value.
	
\begin{figure}[!h]
		\centering
		\includegraphics[scale=.49]{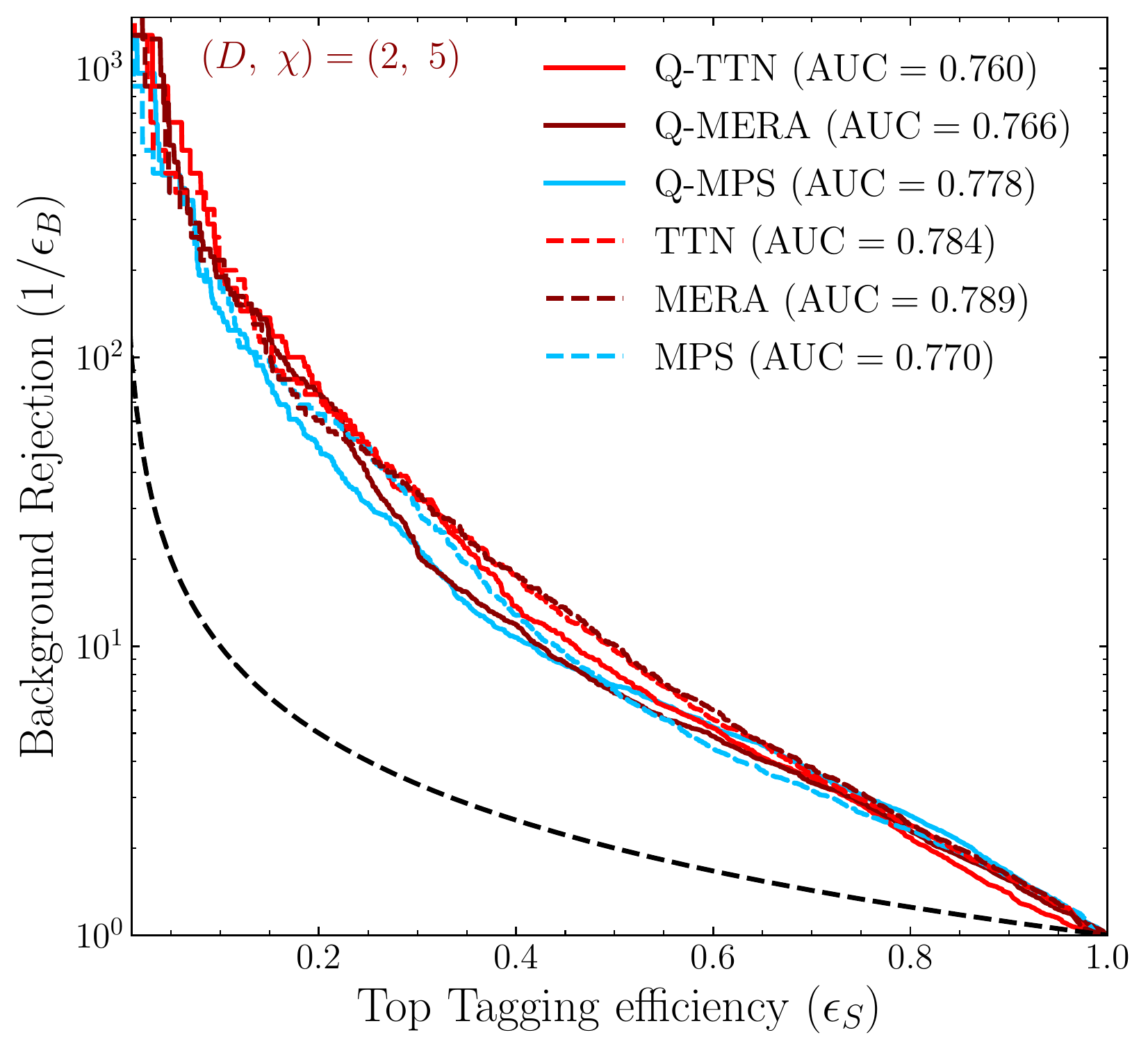}
		\includegraphics[scale=.49]{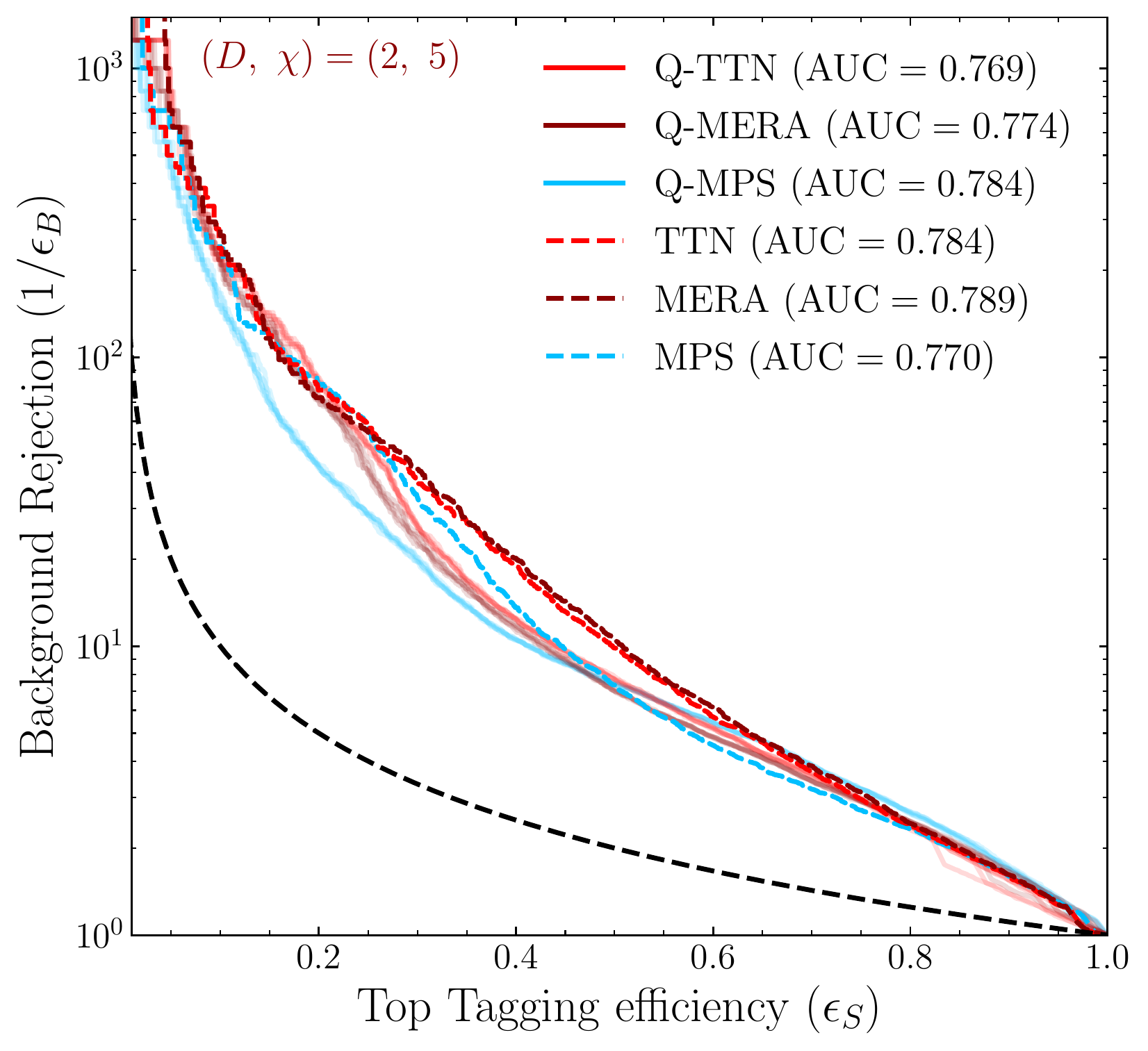}
		\caption{\it ROC curve for four qubit analysis where TTN, MERA and MPS are shown with red, dark red and light blue, respectively. Both plots compare classical TNs with quantum versions where QTNs are shown with a solid, dash-dot and dotted line for TTN, MERA and MPS, respectively, where TNs are shown with dashed lines. The top panel shows the comparison where the QTNs are executed in IBM Quantum hardware (\texttt{ibmq\_quito}), where the bottom panel shows the same for AER simulation. Multiple shaded lines for AER simulation represent different noise model assumptions. Each sample includes the same randomly chosen 10,000 events from the test sample, and the black dashed line shows a random guess. \label{fig:four_qubits}}
	\end{figure}

	Fig.~\ref{fig:four_qubits} shows a receiver operating characteristic (ROC) curve for four qubit test results where we only used 10,000 events from the test set due to hardware limitations. For the comparison with QTN, the bond dimension of each TN ansatz has been chosen to be five, and the Hilbert space dimensions are set to two. Although each wire carries a two-dimensional vector, we observed that $\chi=2$ led to significantly worse results than QTN's for any given TN ansatz. The top panel in Fig.~\ref{fig:four_qubits} shows the results that are executed with the quantum hardware, and the bottom panel shows the quantum simulation execution with various noise models based on five different hardware configurations within IBM Quantum\footnote{Quantum simulations are based on \textsc{Qiskit}'s AER package. The simulated noise models are based on \texttt{ibmq\_belem}, \texttt{ibmq\_bogota}, \texttt{ibmq\_lima}, \texttt{ibmq\_manila}, and \texttt{ibmq\_quito} processors.}. The same colour has visualised the results from each noise model with shaded lines for each ansatz. Each QTN has been executed 5000 times (so-called shots), where the final result is chosen to be the mean of 5000 execution. In both panels, TTN, MERA and MPS have been represented with red, dark red and light blue lines, where QTNs are shown with solid, dotted and dash-dot, and TNs have been shown with dashed lines. For the given four qubit configuration with $ (D, \chi)=(2,5)$ TTN, MPS, and MERA have 90, 130 and 250 trainable parameters, with their quantum counterparts 6, 6 and 8 trainable parameters, respectively. We have also investigated more generic gate configurations with $U(\theta, \varphi, \lambda)$; however, we did not observe a significant improvement in the results. We observe that both TN and QTN lead to similar results for each other. In addition to the ROC curve, each ansatz has been presented with the area under the curve (AUC) value. For both types of ansatz, we observe minimal change between corresponding realisations. This shows that for a small network configuration, $ (D, \chi)=(2,5)$ is sufficient to represent a similar entanglement structure as QTNs.

\begin{figure}[!h]
		\centering
		\includegraphics[scale=.49]{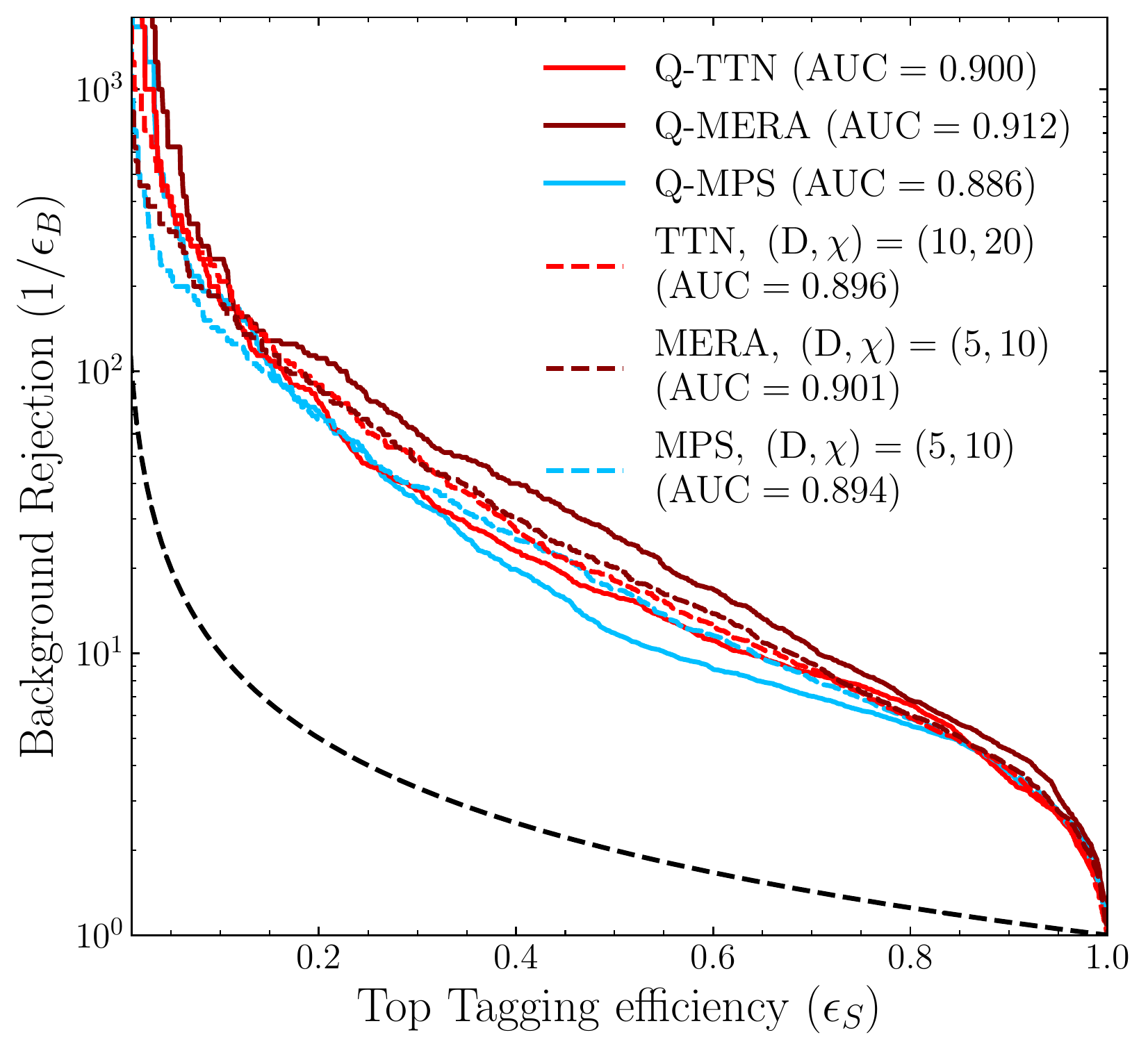}
		\includegraphics[scale=.49]{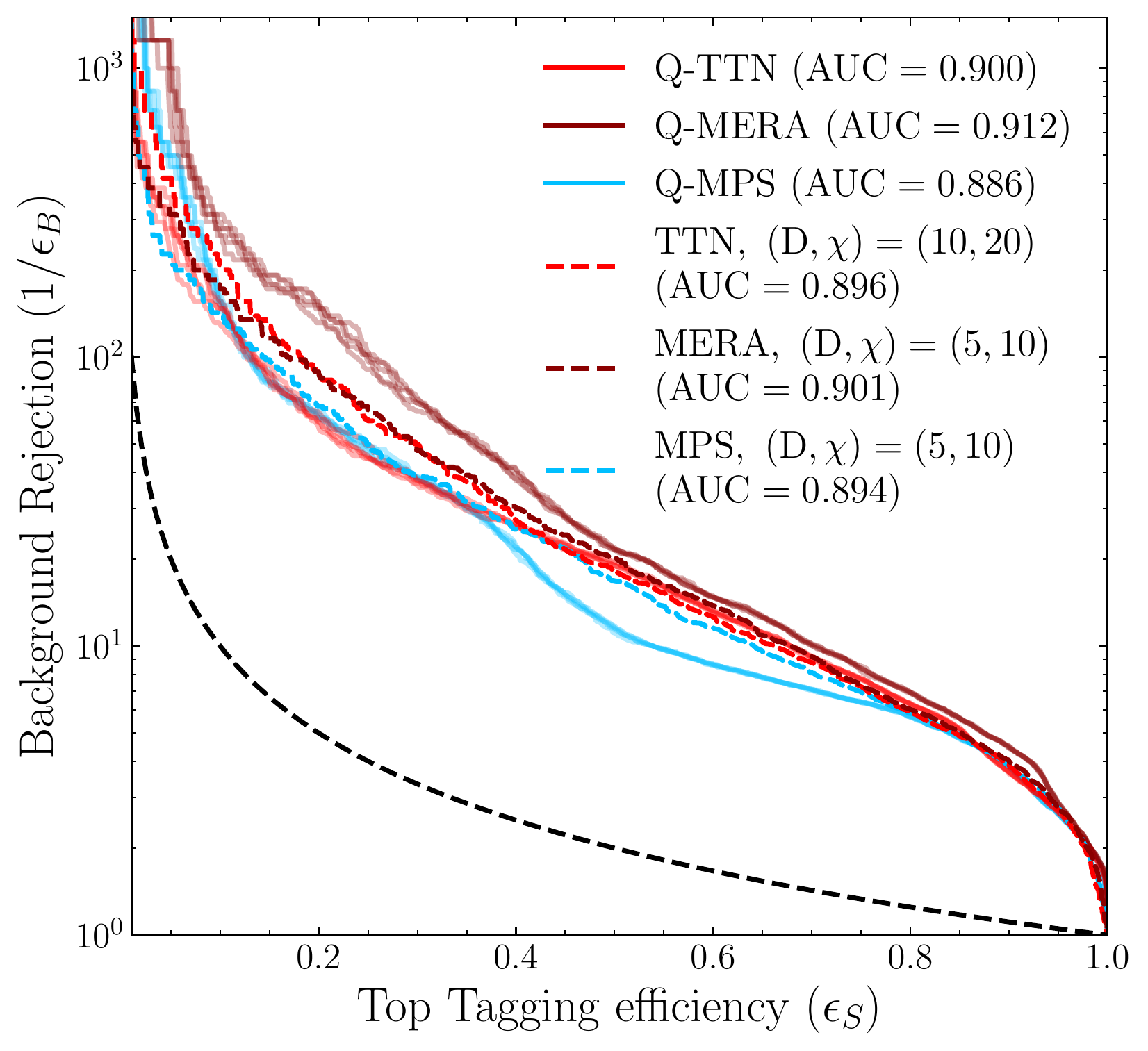}
		\caption{\it Same as Fig.~\ref{fig:four_qubits}, shows this time six qubit arrangement. The top panel shows the samples executed in IBM Quantum hardware (\texttt{ibmq\_perth}).} \label{fig:six_qubits}
	\end{figure}

	Similarly, Fig.~\ref{fig:six_qubits} shows the ROC curve for a six-qubit configuration with the same colour scheme. Again, the top panel shows the results executed with quantum hardware, and the bottom panel shows the quantum simulation results with five different noise models. For both Quantum hardware and simulation, we observed that QTN performance had increased by around 17\%. Although all QTN architectures lead to similar results for a relatively smaller feature space, we observed significant changes between different QTN realizations after adding two more qubits. While for the four qubit configuration, the maximum difference between AUC values of QTNs was 2.4\%, this increased to 3.2\% for the six qubit configuration. Whilst $ (D, \chi)=(2,5)$ leads to comparable results between TN and QTN architectures for four qubit configurations; we observed that once the network size is increased, with such low bond dimension, TNs were not able to reproduce the same performance as QTNs. For this reason, Fig.~\ref{fig:six_qubits} shows TN ansatz with larger $D$ and $\chi$ values where we managed to match the QTN performance by employing $(D,\chi)=(10, 20)$ for TTN and $(D,\chi)=(5,10)$ for MPS and MERA architectures. 

	As a comparison, \autoref{tab:comparisson} shows the values for AUC and the number of trainable parameters for each TN ansatz with different $D$ and $\chi$ values. We observed that, especially for MPS and MERA, only increasing the bond dimension between nodes does not necessarily improve the performance; on the contrary, it has slowed down the optimization process by exponentially increasing the complexity of the loss landscape. Hence increasing Hilbert space dimensions along with the bond dimensions has been shown to perform better with relatively less trainable parameters. Although we did not observe any overtraining for presented values, either method leads to considerable growth in the computational cost of the TN contraction, hence resulting in an inefficient network for ML applications.
	\begin{table}[!h]
		\centering
		\renewcommand{\arraystretch}{1.}
		\begin{tabular}{l|cccc}
			Ansatz & $ D $ & $ \chi $ & \# Parameters & AUC \\\hline\hline
			\multirow{5}{1cm}{TTN}     &  2      & 5 & 235 & 0.755 \\
			        						      &  2      &  10 & 1320 & 0.803 \\
			        						      &  2      &  20 &  9040 & 0.849 \\
			        						      &  5      &  10 & 1950 & 0.873 \\
			        						      &  10    &  20 & 14800 & 0.896 \\\hline
			\multirow{4}{1cm}{MPS}    &  2     & 5 & 230 & 0.811 \\
											      &  2     & 10 & 860 & 0.819 \\ 
											      &  2     & 20 & 3320 &  0.818 \\ 
											      &  5     & 10 & 2150 & 0.894 \\ \hline
			\multirow{4}{1cm}{MERA}  &  2     & 5 & 1225 & 0.850 \\
											      &  2     & 10 & 13400 &  0.840 \\
											      &  2     & 20 & 181600  & 0.848  \\
											      &  5     & 10 & 18200 & 0.901 \\\hline\hline
		Q-TTN & 2 & 2 & 9 & 0.893 \\
		Q-MPS & 2 & 2 & 9 & 0.886 \\
		Q-MERA & 2 & 2 & 17 & 0.914 \\
		\hline
		\end{tabular}
		\caption{\it The upper panel of the table shows AUC values and number of trainable parameters for various Hilbert-space ($ D $) and auxiliary ($ \chi $) dimensions presented for classical TNs. The bottom panel shows the number of trainable parameters and AUC values for the corresponding QTN ansatz. Each ansatz is designed to receive six pixels (qubits); hence the table provides complementary results for Fig.~\ref{fig:six_qubits}.  \label{tab:comparisson}}
	\end{table}


	Although pruned experiments show significant improvement in representability, it is essential to look into a more realistic investigation. For this reason, we employed the full $4\times 4$ image and tested the quality of the classification. Both quantum and classical TNs are reconstructed by simply extending the representations presented above. Instead of using the complete data, however, this time, we used only $10,000$ training events for quantum TNs and $50,000$ training events for classical networks, which gives a significant disadvantage to the quantum TNs. The rest of the training hyper-parameters remained the same as above. Finally, each network has been tested with the same $10,000$ test sample as before. For the classical TNs, we used 10D Hilbert space mapping along with 20 bond dimensions, resulting in $64,800$ parameters for TTN, $56,600$ parameters for MPS and $10,248,004$ parameters for MERA. For their quantum counterparts, on the other hand, we get 30 parameters for QTTN, 30 parameters for QMPS and 60 parameters for QMERA.

\begin{figure}[!h]
	\centering
	\includegraphics[scale=.55]{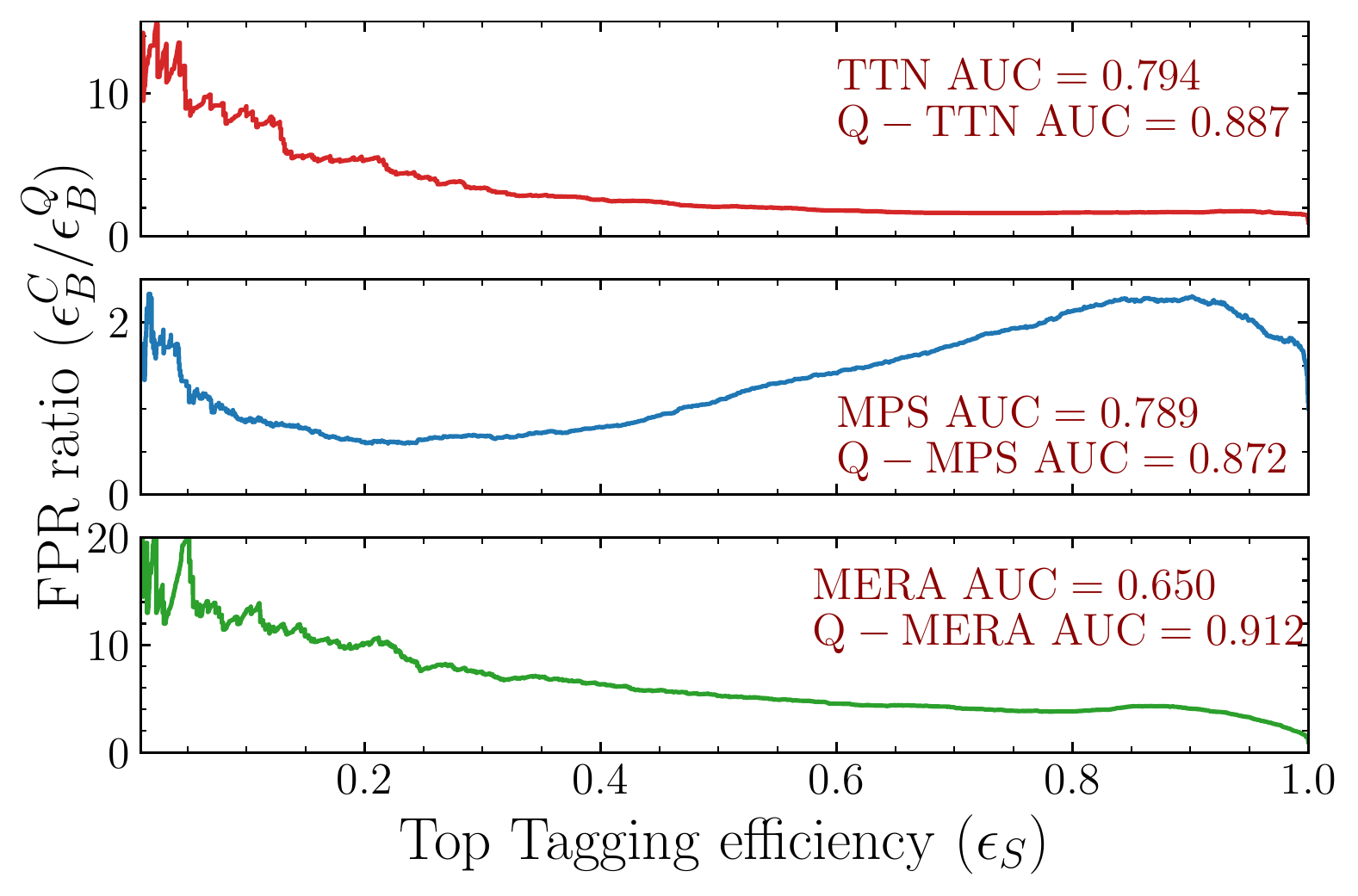}
	\caption{\it Background rejection ratio between quantum and classical ROC curves has been presented for 16 feature analysis. From top to bottom, panels show TTN, MPS and MERA architectures and each panel is presented with a respective AUC value. $(D,\chi) = (10,20)$ has been used for classical networks.} \label{fig:sixteen_qubits}
\end{figure}

	Fig. \ref{fig:sixteen_qubits} shows the False Positive Rate (FPR) ratio between classical ($\epsilon^C_B$) and quantum ($\epsilon^Q_B$) networks, plotted against tagging efficiency. From top to bottom, panels show the ratio curve for TTN, MPS and MERA networks alongside their respective AUC value. For each case, we observe that the performance of the QTNs is much superior to the CTNs. It is also essential to note that when trained with only $10,000$ sample, CTNs performed significantly worse than the presented outcome. We also observed particularly low variance in the gradients of CTNs, which made the training harder. As before, QMERA showed the best performance among the quantum networks, reaching $0.912$ AUC value despite the lack of training samples. Among the classical networks, MPS seems to achieve the closest result to its quantum counterpart; however, as before, all the classical networks require much larger Hilbert space mapping and bond dimensions.
	
	It is essential to emphasise that one might naively assume that the claim of fig.~\ref{fig:sixteen_qubits} is that the QTN accuracy is higher than its classical counterpart. As noted above, we limited our networks with 10D Hilbert space mapping with 20 auxiliary dimensions for classical TNs; this can easily be increased to match the accuracy of the QTNs. However, we have limited our network since it is already clear that classical TNs will require significantly larger Hilbert space mapping and auxiliary dimensions. We also note that we did not observe overtraining; hence it is possible to increase the number of parameters for classical TNs.

\begin{figure*}
		\centering
		\includegraphics[scale=.48]{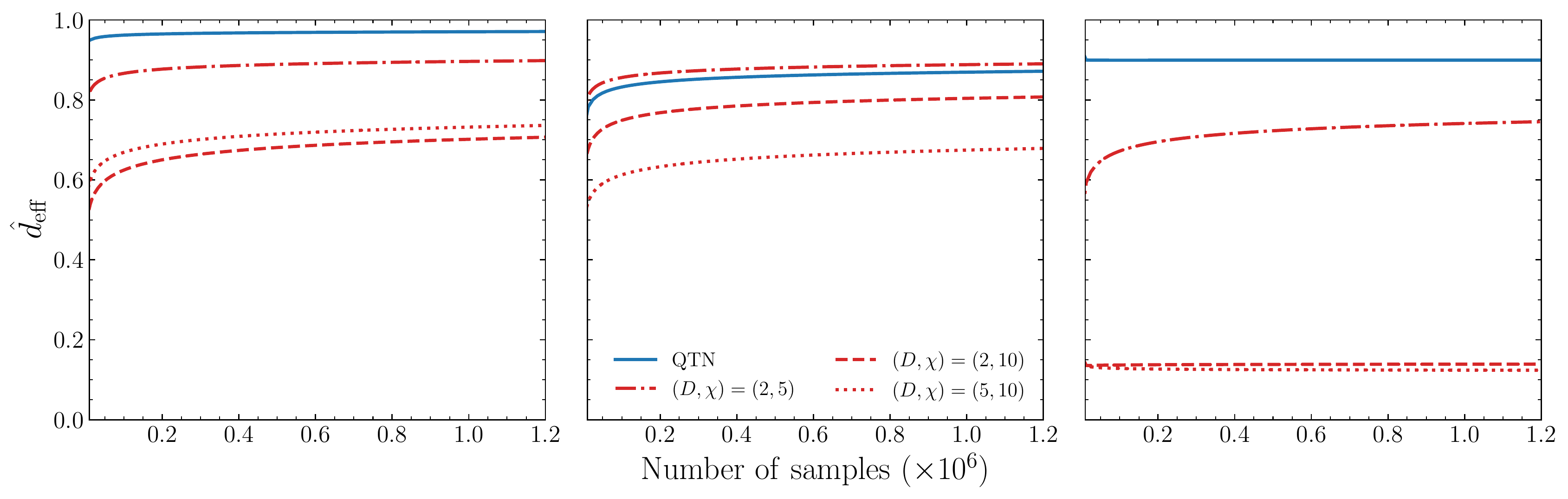}
		\includegraphics[scale=.42]{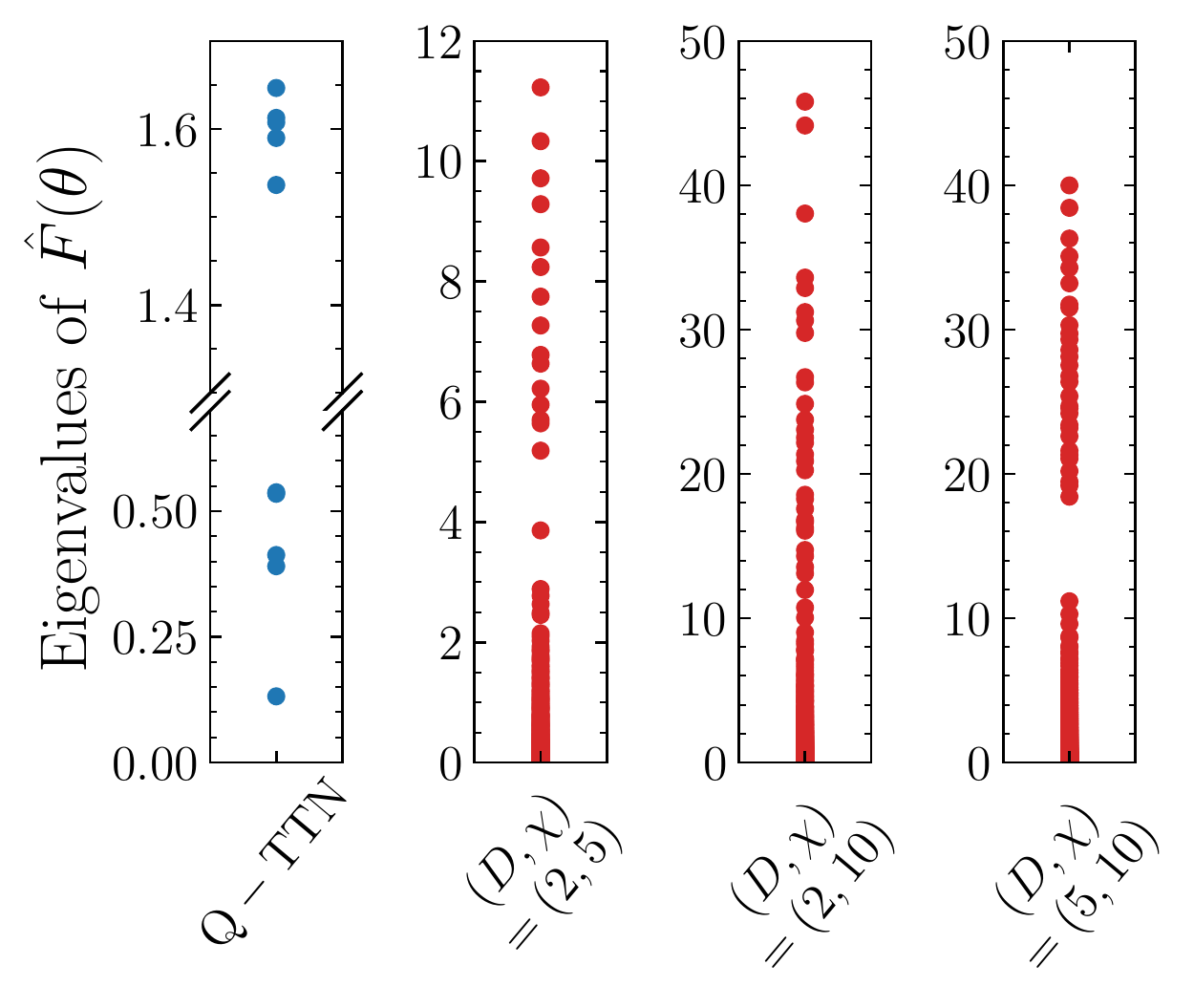}\quad
		\includegraphics[scale=.42]{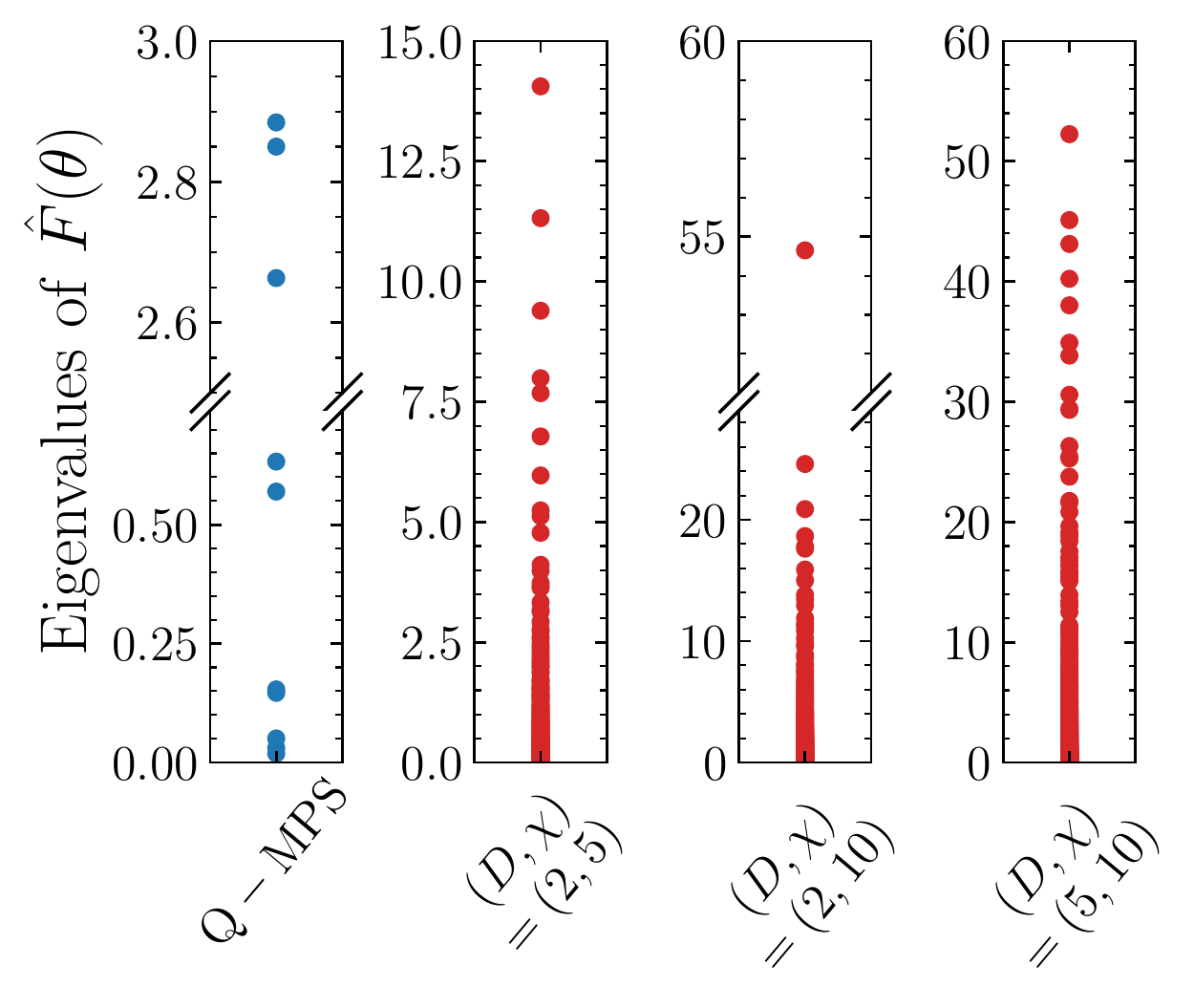}\quad
		\includegraphics[scale=.42]{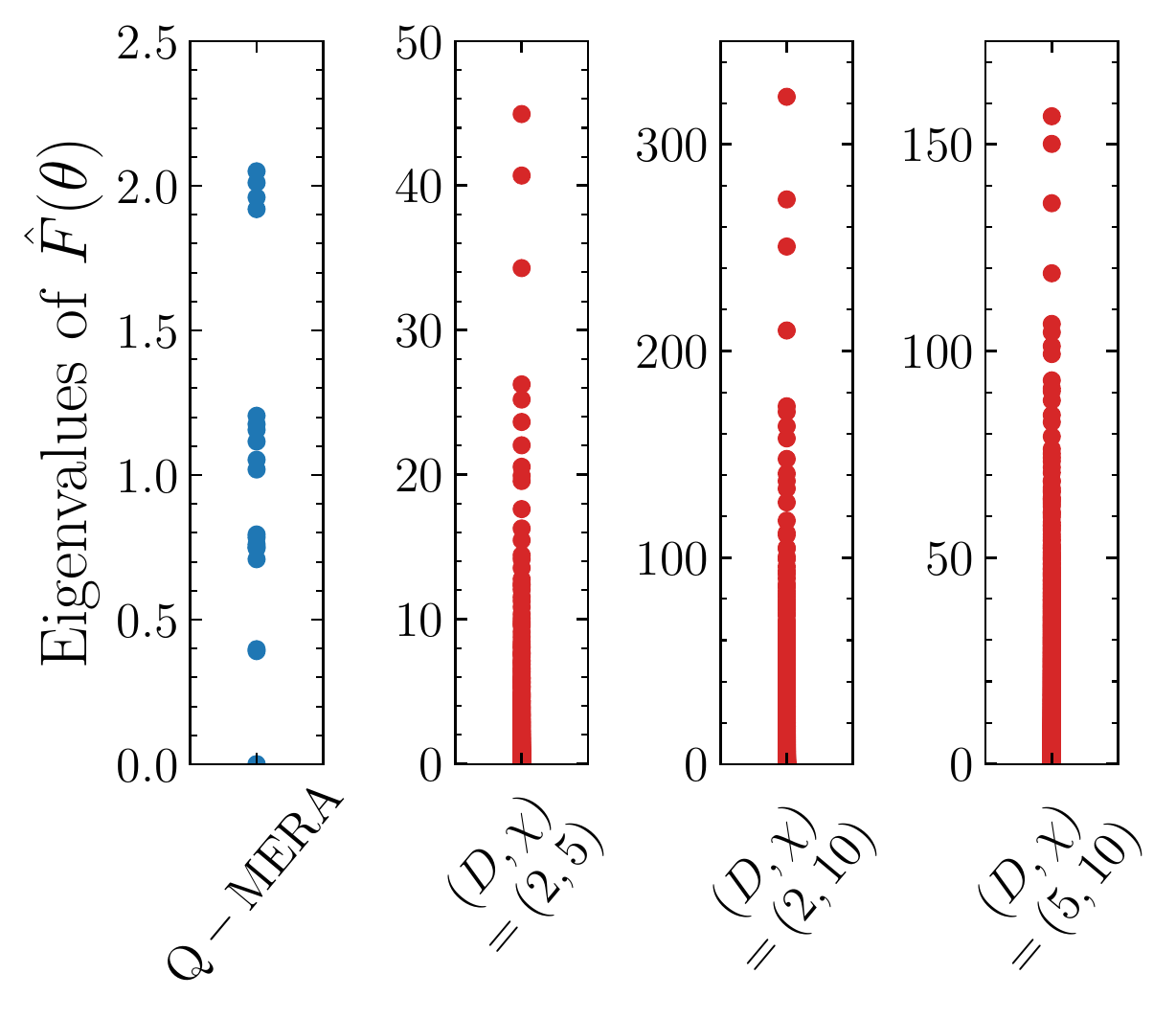}
		\caption{\it Upper panel shows normalised effective dimension with respect to the number of samples for each TN realisation, namely TTN, MPS and MERA, respectively. QTNs are shown with the solid blue line, and TNs shown with dot-dashed, dashed and dotted red lines with respect to corresponding Hilbert space and bond dimensions shown with $ (D,\chi) $.  The bottom panel shows the eigenvalue distributions of the normalised Fisher matrix for each TN realisation with the same colour scheme. Again, the plot flow follows from left to right TTN, MPS and MERA, respectively.} \label{fig:fisher}
	\end{figure*}

	In order to study the increase in the complexity of the optimisation landscape, we employed the Fisher information matrix presented in \autoref{sec:qtn_intro}. The lower three panels of Fig.~\ref{fig:fisher} show the eigenvalue distribution of the normalised Fisher matrix for TTN (left), MPS (middle) and MERA (right). The mean Fisher matrix has been calculated as the average of thousand executions of randomly chosen input and parameter space. Input values have been uniformly varied within $[0,\pi]$ where trainable parameters have been allowed to run within the $[-\pi,\pi]$ range for each ansatz. Each group shows a blue scatter plot for QTNs and red for different $(D, \chi)$ configurations of TNs. We observe that eigenvalue distributions for QTNs are relatively evenly distributed for each case. Q-MPS shows the most uneven distribution among all other QTN configurations, with more eigenvalues around zero. The eigenvalue distribution for TNs, on the other hand, has been observed to have the typical behaviour of a classical network, where the majority of the points are clustered very close to zero, followed by substantially large values. 
	
	The effect of the eigenvalue distribution has been reflected in the normalised effective dimension distribution shown on the upper panel of Fig.~\ref{fig:fisher} following the same colour scheme, where $(D, \chi)$ configuration has been distinguished by using different line styles. Recall that the normalised effective dimension distribution for each network realisation will converge to 1 if a large enough sample has been introduced during the training. We observe that Q-TTN obtains the most significant effective dimension value, whereas the rest has a more prolonged convergence rate. The effect of eigenvalue distribution can directly be observed in the effective dimension distribution; due to the larger cluster around zero, Q-MPS shows the lowest values for the effective dimensions. 

	Whilst a larger TN-bond dimension is essential to have a more accurate representation of the data, we observe that the increase in the bond dimension directly affects the efficiency of gradient-based methods to train the system. The dot-dashed and the dashed lines in Fig.~\ref{fig:fisher} shows $(D,\chi) = (2,5)$ and $(2,10)$. We observe a significant decrease in the effective dimension's convergence rate for each architecture. Similarly, increasing the bond dimension results in a more uneven distribution in the eigenvalues of the Fisher matrix. Comparing these results with \autoref{tab:comparisson}, only increasing the bond dimensions does not provide a sufficient increase in the performance of the network, which also sacrifices the trainability of the ansatz. Increasing the Hilbert space dimension along with the bond dimensions, on the other hand, has been shown to achieve significantly better performance. \autoref{tab:comparisson} shows that by only increasing the Hilbert space dimensions, one can gain around an 8\% increase in the AUC values. However, this plummeted the effective dimension convergence rate, leading to a flat optimisation landscape.

	Despite the significant improvement in the performance, due to the hardware limitations in near-term quantum devices, it is impossible to input a larger $\eta-\phi$ plane into a quantum circuit. However, as shown before, since TNs can effectively represent a quantum many-body system instead of manipulating the phase space to be suitable for a small quantum circuit, TNs can be used as a data processing layer which can then be deposited into a quantum circuit for classification. In the following section, such hybrid architectures will be investigated.

	\subsubsection{Hybrid Quantum -- Classical Tensor Networks}\label{sec:hybrid}

	This section introduces two main types of hybrid classical-quantum TN architectures. Each architecture is designed to have one classical layer, which will process a large image and reduce it to a four-qubit output. A quantum circuit will classify the processed output from the TN in the following. Since the calorimeter images used in this study impose two-dimensional correlations between each pixel, we will adjust each given architecture to capture these correlations effectively. We have constrained the architectures to two main groups, namely one with TTN and another with MPS classical layer. Due to the computational complexity, we did not introduce a classical MERA layer that can process the image within a hybrid ansatz\footnote{An application of MERA on a two-dimensional lattice can be found in ref.~\cite{PhysRevLett.100.240603}.}. 
	
	\begin{figure}[!h]
		\centering
		\includegraphics[scale=1]{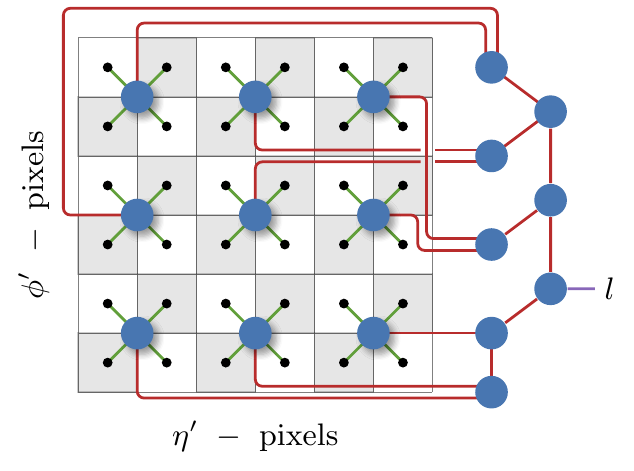}
		\includegraphics[scale=1]{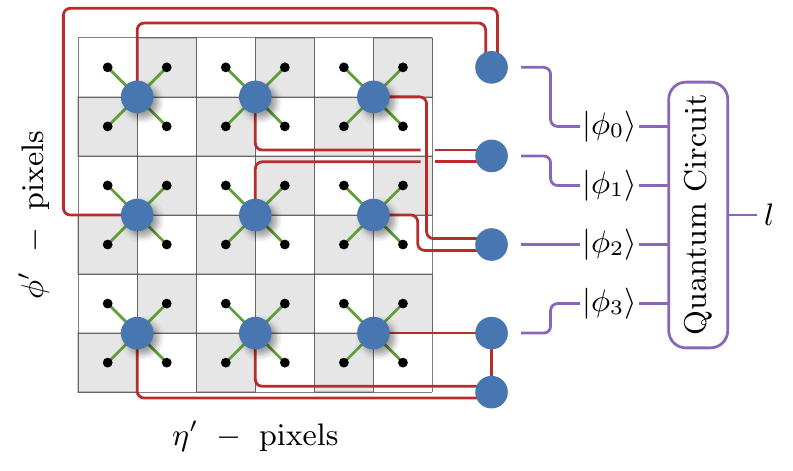}
		\caption{\it Left panel shows the representation of the classical TTN designed to capture 2D mapping. The right panel shows the representation of the hybrid classical--quantum TN design where the classical portion is TTN, and the quantum circuit can be any four-qubit variational circuit. Small black circles represent the pixel values, and green lines correspond to Hilbert space dimensions. Blue circles represent tensor nodes, and the red connections are auxiliary dimensions. Finally, $ l $ stands for the network decision. \label{fig:hybrid_ttn}}
	\end{figure}
	
	As mentioned before, TTNs can capture beyond one-dimensional correlations between lattice sites. Concretely, with a more complex TTN tensor structure, it is possible to capture two-dimensional correlations between pixels of an image~\cite{ttn2019}. We designed a TTN with two types of condenser tensors to achieve this. The tensors connected directly to the image pixels are designed to pool four neighbouring pixels, capturing the correlations in both $\eta$ and $\phi$ axes. The collection of such tensors can be interpreted as a trainable pooling layer where each node takes four pixels and maps them to a vector on a higher-dimensional manifold. Hence each of these nodes forms rank-5 tensors. In the following, each of these nodes is connected strategically through rank-3 tensors until we get the desired dimensionality. For complete classical TTN, the dimensionality is hierarchically reduced until we reach the output dimension. The hybrid realisation has been reduced to a concatenated four-dimensional vector for four qubit input. Fig.~\ref{fig:hybrid_ttn} shows the representation of these two architectures where the network on the top shows pure classical TN, and the bottom panel shows the hybrid version of the architecture. This specific architecture aims to group the most related set of tensors to pool the local correlation information before investigating a more global picture. As before, red and green lines represent auxiliary and physical dimensions. The grid represents the image shown in Fig.~\ref{fig:sig_bkg_im} where the black dots in the centre of each pixel represents the mapped pixel vector defined in Eqs.~\eqref{eq:feature_map} and \eqref{eq:hypersphere}. Purple lines on the bottom panel show the collection of the output values from the classical layer, where each node returns a one-dimensional vector which is then concatenated into a four-dimensional vector to be processed in the quantum circuit. As before, $ |\phi_i\rangle $ indicates valid data embedding into the quantum circuit.

	\begin{figure}[!h]
	\centering
	\includegraphics[scale=1]{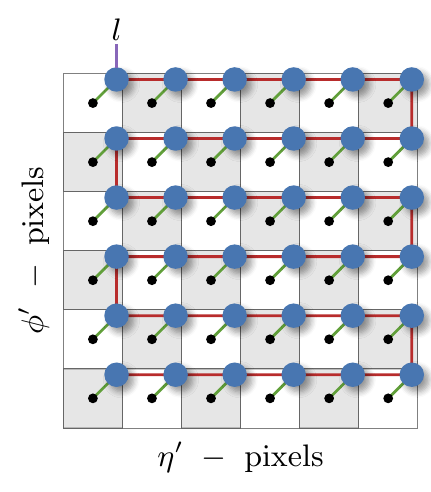}\quad\quad\quad
	\includegraphics[scale=1]{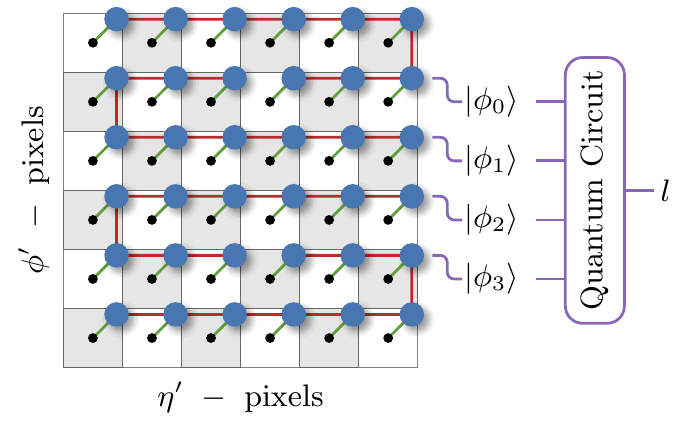}
	\caption{\it Representation of hybrid classical-quantum TNs with MPS as a classical layer. Image has been divided into four blocks of nine nodes of MPS, where each MPS output goes into the quantum circuit. As in Fig.~\ref{fig:hybrid_ttn}, green shows Hilbert dimensions and red shows the auxiliary dimensions. \label{fig:hybrid_mps}}
	\end{figure}

	Due to the nature of MPS, its structure is purely limited to one-dimensional lattices. However, as shown in ref.~\cite{Araz:2021un}, the image can be reshaped so that the locally correlated pixels are close to each other. An s-shaped reshaping procedure is based on the $\eta$--axis. Hence, for the classical MPS, we will strictly follow the previously proposed procedure where the pixels are reordered in the $\eta$-based s-shaped reordering procedure. For hybrid architecture, on the other hand, the MPS chain has been divided into four blocks of nine nodes where each outputs a one-dimensional vector which then is concatenated and inputted into a quantum circuit. Fig.~\ref{fig:hybrid_mps} shows the representation of these architectures following the same colour scheme as before. The top panel represents the classical MPS, and the bottom panel shows the hybrid architecture. 

	Such hybrid architecture poses the question of how to train such a network. Although SGD has been proven to be a highly effective training method, it has been repeatedly shown that the QNGD method can achieve much faster convergence for a quantum circuit. Hence we used a mixed optimization algorithm where the classical portion is trained with the \texttt{Adam} algorithm with an initial learning rate of $10^{-4}$ and the quantum circuit has been optimized via QNGD with an initial learning rate of $10^{-2}$. Both learning rates are decayed with the same factor simultaneously, whereas if the loss value of the validation set did not improve for 25 epochs, the learning rate has been reduced by a factor of 0.5. All ansatz are trained with the complete training sample with a batch size of 100 events. 

	\begin{figure}[!h]
	\centering
	\includegraphics[scale=.49]{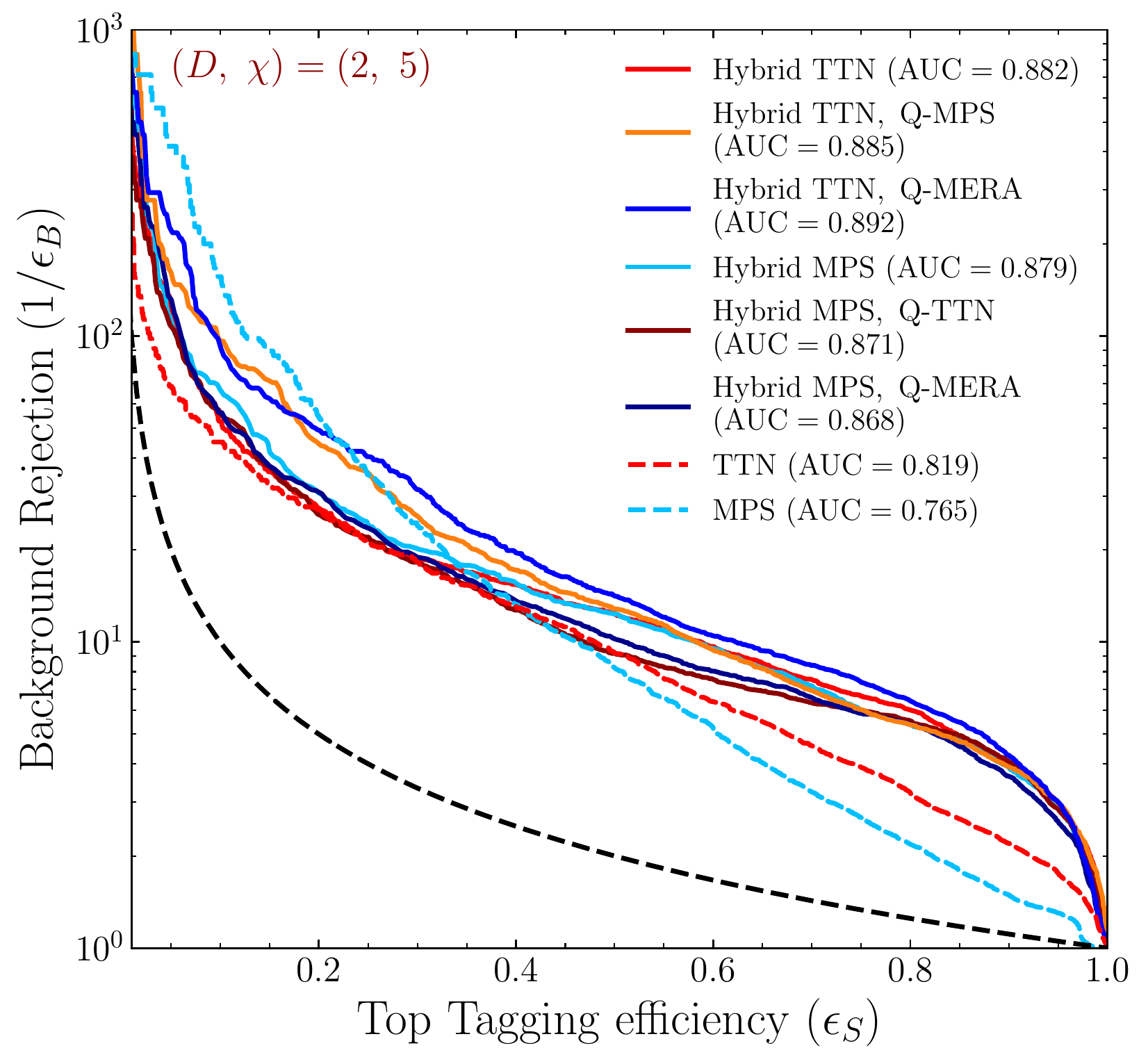}
	\includegraphics[scale=.49]{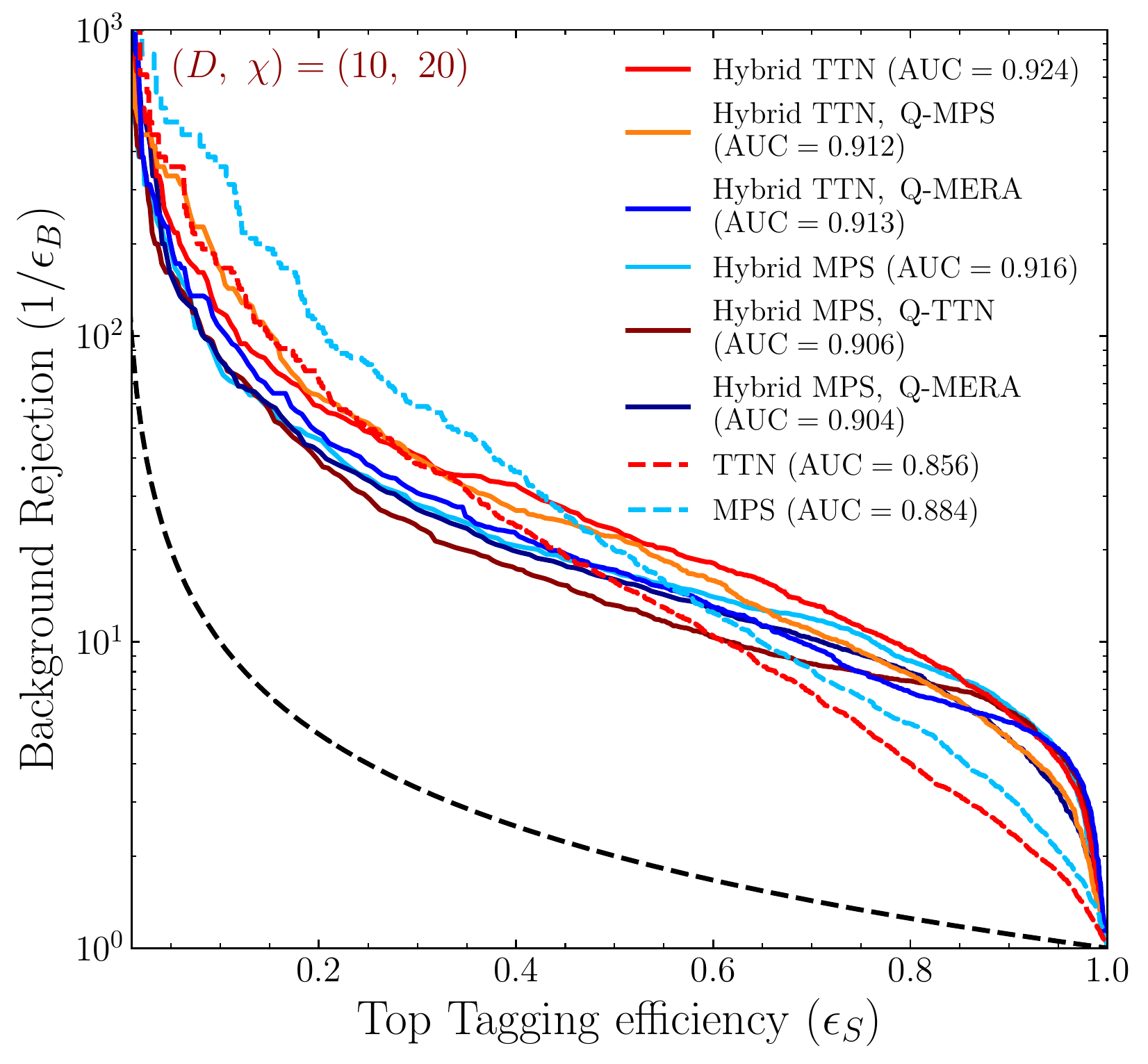}
	\caption{\it Same as Fig.~\ref{fig:four_qubits}, shows this time hybrid classical-quantum TN ansatz with four qubit arrangement. All quantum circuits have been executed in IBM Quantum hardware (\texttt{ibmq\_quito}). \label{fig:hybrid_roc}}
	\end{figure}

	Fig.~\ref{fig:hybrid_roc} presents the results of various realisations of the hybrid TN architectures compared to purely classical counterparts, where the top panel shows the results generated by setting $ (D, \chi)=(2,5)$. The bottom panel shows the same for $ (D, \chi)=(10,20)$. The dashed red and light blue curves represent both panels' purely classical TTN and MPS realisations. Solid curves represent hybrid ansatz with different QTNs. The architectures with TTN are shown in red, orange and dark blue, representing the networks with Q-TTN, Q-MPS and Q-MERA. The light blue, dark red and dark blue curves, on the other hand, show the MPS with Q-MPS, Q-TTN and Q-MERA, respectively. Due to the effective two-dimensional representation in TTN, we observe a 7\% increase in the performance compared to MPS for $ (D, \chi)=(2,5)$ configuration. This performance increase mainly originated from the high-efficiency regime, $\epsilon_S \gtrsim 0.4$, but MPS performs better in the low-efficiency. MPS and TTN possess 1730 and 1645 parameters in this configuration, respectively. However, due to the growth in network complexity, this changes for $ (D, \chi)=(10,20)$ configuration where MPS and TTN consist of $ 136,600 $ and $ 1,856,800 $ parameters, respectively. As before, TN's optimisation capabilities degrade with the complexity of the network, where gradient-based methods are insufficient to train the network efficiently. Hence, although TTN architecture can interpret 2D objects more efficiently than MPS, it performs worse due to the complexity of the network by a factor of 3\%. 

	We implemented a hybrid test for each classical TN layer using all four qubit QTNs mentioned in previous sections. As shown with the solid lines, each hybrid realisation performs better than the classical versions. However, this improvement is solely based on the high-efficiency regime for each configuration. This might be due to the sensitivity of the networks to the subtle information on the image. The QCD background is mainly concentrated on a few pixels; hence, the information can quickly vanish in a complex network structure if the impact is not significant enough. Whilst for $ (D, \chi)=(2,5)$ configuration, we do not observe a large difference in performance for different hybrid models, the difference gets larger with $ (D, \chi)=(10,20)$ configuration. Since we used the same number of qubits with the same QTNs, we conclude that this is due to the increase in the mapping capability of the classical layer. For each configuration, the TTN layer has been observed to have a larger impact on the correct classification of the data. However, for the more extensive configuration, hybrid ansatz with the MPS layer has been observed to achieve much closer results to TTN.

	\section{Conclusion}\label{sec:conclusion}

	Tensor Networks are algebraic tools to represent high-rank tensors effectively. They have been widely studied to represent complex quantum many-body systems and capture their entanglement properties. Multimodal data can be expressed as a quantum system, and TNs can effectively represent correlations within the data structure. Moreover, due to the ability to describe quantum states, TNs are the ideal machinery to study quantum machine learning, where the expertise on ``classical'' methods built for TNs can directly be applied to quantum hardware.

	In this study, we explored the possibility of classifying HEP data with TN-inspired quantum circuits. We mainly focused on three widely studied TN architectures: TTN, MPS, and MERA and compared their performance to corresponding quantum circuits. We have shown that, although classical TNs are very successful in representing complex data structures, they require large auxiliary and Hilbert space dimensions to capture the natural entanglement capabilities of a quantum system. Hence, to achieve the same performance as QTNs, TNs must be executed with a much higher computational cost and more trainable parameters. We additionally observed that TNs require much more training data compared to their quantum counterpart, to be able to sufficiently generalise the output.
	
	The Fisher information matrix provides a Riemannian metric to measure the flatness of the optimization landscape. Based on Fisher information, the effective dimensions relate to the number of samples required to represent the statistical model well. Whilst increasing the network's dimensionality helped improve the performance, using the Fisher information, we observed that this resulted in exponentially suppressed gradients, rendering gradient-based methods unable to train classical TNs efficiently. Additionally, using effective dimensions, we show that TNs require exponentially more data to achieve sufficient representation of the data with increasing auxiliary and Hilbert space dimensions. Thus, we find that QTNs can perform significantly better than TNs with a fraction of trainable parameters.

	Despite the undeniable success of QTNs, they are still constrained to a low number of qubits due to the limitations of quantum hardware in near-term quantum devices, which results in the inability to learn the entire dataset. To surpass this limitation, we proposed a hybrid end-to-end training architecture where a larger phase space of data can be processed with classical TNs and then deposited into a QTN to be classified. Due to the nature of the TNs, this allows a flexible architecture. As a result, more classical nodes can be transformed into circuit inputs once more qubits are available. Finally, we compared the purely classical TNs with hybrid architectures and showed that hybrid networks perform much better than strictly classical TNs.
	
	In this study, we have limited our QTN architecture to a modest size where each circuit block is transformed into two input states. Although such simple architecture already showed the quantum advantage over classical TNs, these can be extended to blocks transforming more qubits. Additionally, the entanglement between blocks can be enhanced by introducing auxiliary qubits to increase the correlations between circuit blocks. This can significantly improve the expressivity of the quantum network. Additionally, the hybrid architectures presented in this study were highly simplistic. Since specific geometric properties can be embedded into TN architecture, much more complex structures can be employed using the known symmetries within the data.

	\section{Acknowledgements}
	We acknowledge the use of IBM Quantum services for this work. The views expressed are those of the authors, and do not reflect the official policy or position of IBM or the IBM Quantum team. In this paper we used \texttt{ibmq\_quito} and \texttt{ibm\_perth}, which are one of the IBM Quantum Falcon Processors. We thank Vishal S. Ngairangbam and Josh Izaac for very helpful discussions.

	\bibliography{bibliography}

\end{document}